\documentclass[twocolumn]{aa}
\usepackage{graphicx,epsfig}
\begin{document}
   \title{Deep Extragalactic VLBI-Optical Survey (DEVOS)}

   \subtitle{I. Pilot MERLIN and VLBI observations}

   \titlerunning{Deep Extragalactic VLBI-Optical Survey (DEVOS) I.}

   \author{L. Mosoni\inst{1,2}, S. Frey\inst{2}
           \and
           L.I. Gurvits\inst{3}, M.A. Garrett\inst{3}
           \and
           S.T. Garrington\inst{4} 
           \and
           Z.I. Tsvetanov\inst{5}
          }

   \offprints{L\'aszl\'o Mosoni, e-mail: {\tt mosoni@konkoly.hu}}

   \authorrunning{L. Mosoni et al.}

   \institute{Konkoly Observatory of the Hungarian Academy of Sciences, P.O. Box 67, H-1525, Budapest, Hungary
             \and
             F\"OMI Satellite Geodetic Observatory, P.O. Box 585, H-1592, Budapest, Hungary
               \and
              Joint Institute for VLBI in Europe, Postbus 2, 7990 AA Dwingeloo,The Netherlands
              \and
              University of Manchester, Jodrell Bank Observatory, Macclesfield, Cheshire SK11 9DL, UK
              \and
              Department for Physics and Astronomy, Johns Hopkins University, Baltimore, MD 21218, USA
     }

   \date{Received 19 May 2005; accepted 2 August 2005}

   \abstract{We present the results of the pilot observations of the Deep
Extragalactic VLBI-Optical Survey (DEVOS). Our ultimate aim is to collect 
information on compact structures in a large sample of extragalactic radio 
sources ($\sim10^4$ objects) up to two orders of magnitude fainter than 
those studied in typical imaging Very Long Baseline
Interferometry (VLBI) surveys up until now. This would
lead to an unprecedented data base for various astrophysical, astrometric
and cosmological studies. The first global VLBI observations of the 
DEVOS programme were successfully conducted in May 2002. We selected 
sources without any spectral criterion from the Very Large Array (VLA) 
Faint Images of the Radio Sky at 
Twenty-centimeters (FIRST) catalogue, that are also detected with the
Multi-Element Radio Linked Interferometer Network (MERLIN).
The DEVOS pilot sample sources are in the area of the sky that is 
covered by the Sloan Digital Sky Survey (SDSS).
We describe the sample selection and present high resolution 5-GHz radio
images of the sources. Based on the results of
this pilot study, we estimate the outcome of and the resources needed for
a full-scale DEVOS project.

   }

   \maketitle
%

\section{Introduction}

Very Long Baseline Interferometry (VLBI) imaging surveys provide data
\mbox{--~milli}-arcsecond (mas) resolution \mbox{images~--} for large samples
of extragalactic radio sources, thus enabling studies of the nature of 
physical phenomena in the close vicinity of the central engine in active 
galactic nuclei (AGNs). Completed and ongoing VLBI surveys are reviewed by 
Gurvits (\cite{gurvits02}).

Until recently, large-scale VLBI surveys at GHz frequencies were usually 
targeted, and the samples are flux density limited at $S_{\rm{total}} \gse 100$~mJy. 
The surveys often apply spectral selection criteria to observe 
mostly flat-spectrum radio sources (e.g. Beasley et al. \cite{beasley02}, 
Scott et al. \cite{vsop}). These samples are therefore dominated 
by relatively nearby ($z \lse 1$) luminous AGNs. 
Novel approaches -- aiming at the fainter radio source population 
-- like phase-reference imaging around strong calibrator sources, and deep 
wide-field imaging in particular, are summarised by Garrett et al. 
(\cite{garrett04a}).

In order to reach conclusive results on the intrinsic properties of sources as
well as possible imprints of cosmological models in the source structures, one
has to match in luminosity sources detected and imaged with VLBI at low 
redshift (e.g. $z < 0.2$) with those at high redshift ($z > 1$). 
This requires VLBI study of $z > 1$ sources of
luminosities $10^{23} - 10^{25}$~W~Hz$^{-1}$, corresponding to
mJy-level flux densities.
This work can be carried out with sensitive VLBI
arrays (lower noise, larger apertures, broader bands, higher stability of 
the signal paths) using innovative observing techniques.
It now seems feasible to increase
the total number of VLBI-imaged extragalactic sources to at least $\sim10^4$.

The applications of data from VLBI surveys include astrophysical and
cosmological studies of powerful compact radio sources.
In order to place the objects in the cosmological context,
redshift measurements are also required. With the advent of
large optical surveys (e.g. the Sloan Digital Sky Survey, SDSS, 
Abazajian et al. \cite{abaza03,abaza04}; the 2dF Survey,
Croom et al. \cite{croom}), the problem of missing optical
identifications (morphological type, redshift) will be partly
eliminated within the next couple of years. We can match the sky coverage of
optical and VLBI surveys by selecting sources for VLBI
observations from celestial areas covered by the optical surveys. 

It is hard to predict in detail what kind of scientific insight will become
possible with such a dramatic increase in the number of VLBI-imaged sources.
One can briefly mention a few possible applications:

\begin{itemize}
 \item The {\it cosmological evolution} of radio-loud active galaxy
population could be studied with the data base in the light of
orientation-dependent unification models of powerful radio sources (cf.
Wall \& Jackson \cite{wall}). Note that we do not introduce initial sample
selection criteria based on source spectral index (see Sect.~\ref{sect2}).
 \item {\it Morphological classification} of low-luminosity sources could be
compared with that of their high-luminosity counterparts, in view of possible
evolutionary effects.
 \item The arcsecond-scale and high-resolution radio
morphology of the sources could be investigated to look for signatures of
{\it restarted activity} in AGNs (e.g. Marecki \cite{marecki}).
 \item Sub-samples of {\it less powerful radio sources}, e.g. Compact Steep Spectrum
(CSS) objects could be selected and studied to determine their morphological,
physical and evolutionary properties compared to the well-studied powerful objects
(e.g. Kunert et al. \cite{kunert}).
 \item The data base would be useful to study {\it gravitational lensing}
since the survey of weak compact radio sources probes the parent population
of lensed objects being found in e.g. the Cosmic Lens All-Sky Survey (CLASS)
of $\sim 10^{4}$ weak
flat-spectrum sources (e.g. Myers et al. \cite{myers}).
 \item Several thousand mas-scale images of optically identified extragalactic
sources with known redshifts are needed to bring about conclusive estimates of
fundamental {\it cosmological parameters}, such as the density parameters
$\Omega_{m}$ and $\Omega_{\Lambda}$ (Gurvits \cite{gurvits03} and
references therein; Jackson \cite{jackson}).
As shown by Gurvits (\cite{gurvits94}), such a cosmological test could be
conducted with a very limited VLBI $(u,v)$-coverage using visibility
data, and does not require high dynamic range imaging.
 \item Getting VLBI images of $\sim 10^{4}$ extragalactic sources will be an
essential supplement to and basis for future development of the {\it
astrometric} VLBI data bases (e.g. Johnston et al. \cite{johnston}, Ma et al. \cite{ma}).
The sensitive next-generation space-borne optical astrometry missions (e.g.
Gaia, Perryman et al. \cite{perryman}) would provide a possibility to
directly link the radio and optical reference frames using a large number of
AGNs observed also with VLBI.
 \item Some of the sources detected can serve as in-beam phase-reference 
calibrators for future VLBI observations (Garrett et al. 
\cite{garrett04b}). These observations need accurate positions of the 
sources which can be provided by the DEVOS programme.
\end{itemize}

In this paper we present Multi-Element Radio Linked Interferomerer 
Network (MERLIN) and global VLBI imaging results at 5~GHz of a sample of 
47 radio sources initially selected from the Faint Images of the Radio 
Sky at Twenty-centimeters (FIRST) survey (White et al. \cite{white}). By 
means of this pilot study we assess the feasibility of this approach to 
eventually image thousands of faint radio sources at mas resolution.

The target selection and the
observations are described in Sect.~\ref{sect2}. The data reduction
procedures are presented in Sect.~\ref{sect3}. The results are shown and
discussed in Sect.~\ref{sect4}. A summary and discussion of future 
observations are given in Sect.~\ref{sect5}. The MERLIN and VLBI images 
with the image parameters are presented in the online version of 
this paper only.

\section{Sample selection and observations}
\label{sect2}

The aim of DEVOS programme is to obtain high-resolution radio images of 
compact radio sources considerably weaker than currently available in 
other large-scale imaging VLBI surveys.
The technique of phase-referencing (e.g. Lestrade et al. \cite{lestrade};
Beasley \& Conway \cite{beasley}) is employed in order to
extend the coherence time by using regularly
interleaving observations of a nearby bright and compact calibrator source.
Phase-referencing can be applied for surveying a number of
faint objects in the vicinity of an adequately chosen
reference source. In this case we reverse the usual logic of first selecting the
target object and then looking for a suitable reference source. If a calibrator
is selected first, and the potential targets populate its close vicinity
densely enough, then a single calibrator can serve as phase-reference source
for all of the targets. The different target
sources are observed between the regular intervals spent on the
reference source. Each target source is observed several times during the
experiment which improves the $(u,v)$-coverage and hence
the image quality. The method we applied for our observations was introduced
and demonstrated first by Garrett et al. (\cite{garrett98}) and
Garrington et al. (\cite{gg99}). Wrobel et al.
(\cite{wrobel}) also used this method for investigating parsec-scale radio
properties of about 200 faint FIRST sources. Potential calibrator sources
are relatively common and well distributed (e.g. Beasley et al.
\cite{beasley02}), therefore practically the whole sky north of 
$-30\degr$ declination can be surveyed using this method.

Another aim of our survey is to ensure that many of the sources 
will have optical identifications and redshifts. In particular, this can 
be achieved if the sources are selected from the fields that are being 
surveyed in the SDSS. According to recent studies (Ivezi\'c et al.
\cite{ivezic}), optical counterparts of at
least $30\%$ of all FIRST radio sources are expected to be identified by SDSS.
The SDSS-FIRST catalogue is expected to contain spectra for 
$\sim$15\,000 quasars.

The source sample for the DEVOS pilot observations described here was
defined in 2000, preceding any of the SDSS data releases. 
The bright flat-spectrum radio-loud AGN J1257+3229 was selected 
as calibrator, keeping the potential survey areas of SDSS in mind. 
Optical imaging data of the celestial area surrounding the calibrator
J1257+3229 have recently become available in the latest SDSS Data 
Release\footnote{SDSS Data Release 4 (DR4) -- {\tt http://www.sdss.org/dr4/}}. 
The phase-reference calibrator with a total flux density of 530~mJy 
at 5~GHz is close to the North Galactic Pole (NGP).
The object is also known as an X-ray source identified in the ROSAT
All-Sky Survey (Brinkmann et al. \cite{brinkmann}). Its equatorial
coordinates in the
International Celestial Reference Frame (ICRF) are given in the VLBA
Calibrator Survey\footnote{{\tt http://magnolia.nrao.edu/vlba\_calib/}}
(Beasley et al. \cite{beasley02}):
right ascension $\alpha=12^{\rm{h}}57^{\rm{m}}57\fs231863$,
declination $\delta=+32{\degr}29{\arcmin}29\farcs32604$.
The accuracy is 0.5~mas in both $\alpha$ and $\delta$.

Due to the limited observing resources,
we selected sources that are bright and compact
enough that they might be detectable with VLBI at high angular resolution,
within a reasonable integration time.
Note that no criterion was set based on the spectra: radio sources with
either steep or flat spectra could be selected as well.
The extragalactic radio sources for the DEVOS pilot project described here
were chosen from the FIRST survey data base obtained with the
US National Radio Astronomy Observatory (NRAO) Very Large Array (VLA)
at the frequency of 1.4~GHz (White et al. \cite{white}), 
according to the following selection criteria:

\begin{enumerate}
  \item the integrated flux density at 1.4~GHz is $S_{1.4} > 30$~mJy,
  \item the angular size of the source is $\theta < 5\arcsec$
  (i.e. unresolved with the VLA in the FIRST),
  \item the angular separation of the target source from the
  phase-reference calibrator source selected is less than $2\degr$.
\end{enumerate}

47 sources match the first two criteria from the 1042 FIRST sources found 
within the search radius. The sources potentially
suitable for subsequent VLBI imaging were chosen based on 5-GHz MERLIN
detections of the sample at an angular resolution of $\sim 50$~mas.
The sources detected with MERLIN were then
observed with VLBI. This subsample was imaged at 5~GHz with a 
global VLBI network at $1-3$~mas resolution.

\begin{table*}
 \caption{The 37 sources detected with MERLIN in the DEVOS NGP sample.}
 \label{sample_ordered}
 \begin{tabular}{@{}lllrrrrrrc}
  \hline
  Source name &
\multicolumn{2}{c}{Equatorial coordinates (J2000)} &
FIRST & GB6 &
\multicolumn{2}{c}{MERLIN 5~GHz} &
\multicolumn{2}{c}{VLBI 5~GHz} & Optical ID \\
         & R.A. & Dec. & int & int & peak & int & peak & int \\
  \hline
  & h\hspace{3mm}m\hspace{3mm}s & $\degr$\hspace{4mm}$\arcmin$\hspace{3mm}$\arcsec$ & mJy & mJy & mJy/beam & mJy & mJy/beam & mJy \\
  \hline
  J124833.1+323207 & 12 48 33.12774 & 32 32 07.2818 & 41.91 & --\hspace{1.8mm} & 9.4 \# & 18.3 & 5.2 \# 
& 13.6 \\ 

J124912.4+324438 & 12 49 12.4943 & 32 44 38.765  & 37.46 & --\hspace{1.8mm} & 2.0\hspace{3.8mm} & 3.3 & 
$\leq$ 0.8\hspace{3.8mm} \\ 

J124954.5+332330 & 12 49 54.5037 & 33 23 30.086 & 460.29 & 377\hspace{1.8mm} & 2.2\hspace{3.8mm} & 1.8 & 
$\leq$ 0.8\hspace{3.8mm} \\ 

J125106.9+320906 & 12 51 06.9619 & 32 09 06.392 & 197.27 & 59\hspace{1.8mm} & 5.4 \& & 23.7 & $\leq$ 
0.6\hspace{3.8mm} & & A \\ 

J125214.1+333109 & 12 52 14.18760 & 33 31 09.7061 & 122.64 & 53\hspace{1.8mm} & 10.1 \& & 19.7 & 
1.4\hspace{3.8mm} & $-$  \\ 

J125221.2+313029 & 12 52 21.2646 & 31 30 29.583 & 31.22 & --\hspace{1.8mm} & 2.7\hspace{3.8mm} & 3.4 & 
$\leq$ 0.8\hspace{3.8mm} \\ 

J125240.2+331058 & 12 52 40.28575 & 33 10 58.1931 & 214.18 & 167\hspace{1.8mm} & 129.1 \# & 132.1 & 74.9 
\# & 154.0 & AS \\ 

J125252.8+314419 & 12 52 52.8420 & 31 44 19.502 & 46.19 & --\hspace{1.8mm} & 3.7\hspace{3.8mm} & 5.4 &  
$\leq$ 0.7\hspace{3.8mm} \\ 

J125354.0+311556 & 12 53 54.0410 & 31 15 56.703 & 78.55 & --\hspace{1.8mm} & 2.3\hspace{3.8mm} & 2.0 & 
$\leq$ 0.7\hspace{3.8mm} & & AS \\ 

J125410.8+330912 & 12 54 10.83148 & 33 09 12.0372 & 35.69 & --\hspace{1.8mm} & 7.7\hspace{3.8mm} & 14.6 
& 1.8\hspace{3.8mm} & $-$ & S  \\ 

J125452.4+311819 & 12 54 52.47075 & 31 18 19.8254 & 97.08 & --\hspace{1.8mm} & 10.2 \# & 15.2 & 2.2 \# & 
9.9 \\ 

J125546.2+304611 & 12 55 46.2141 & 30 46 11.461 & 197.96 & $81^{a}$ & 2.7\hspace{3.8mm} & 3.0 & $\leq$ 
0.6\hspace{3.8mm} \\ 

J125620.6+304525 & 12 56 20.6524 & 30 45 25.514 & 32.03 & 20\hspace{1.8mm} & 2.9\hspace{3.8mm} & 6.9 & 
$\leq$ 0.9\hspace{3.8mm} \\ 

J125658.4+325053 & 12 56 58.45917 & 32 50 53.9630 & 205.86 & 54\hspace{1.8mm} & 5.6\hspace{3.8mm} & 17.8 
& 0.9\hspace{3.8mm} & $-$ & S \\ 

J125728.6+342428 & 12 57 28.6610 & 34 24 28.998 & 66.91 & --\hspace{1.8mm} & 4.5\hspace{3.8mm} & 9.4 & 
$\leq$ 1.4\hspace{3.8mm} &  & S \\ 

J125734.2+335801 & 12 57 34.21675 & 33 58 02.0043 & 71.14 & $33^{b}$ & 2.6\hspace{3.8mm} & 11.1 & 
0.6\hspace{3.8mm} & $-$  \\ 

J125734.8+335812 & 12 57 34.8440 & 33 58 12.965 & 40.08 & $33^{b}$ & 2.7\hspace{3.8mm} & 1.9 & $\leq$ 
0.6\hspace{3.8mm} \\ 

J125755.7+313915 & 12 57 55.72881 & 31 39 15.3298 & 32.10 & --\hspace{1.8mm} & 11.6 \#  & 11.7 & 9.0 \# 
& 12.0 & S \\ 

J125842.2+341109 & 12 58 42.20288 & 34 11 09.5319 & 38.12 & --\hspace{1.8mm} & 4.1 \# & 4.2 & 0.79 \# & 
2.9 \\ 

J125858.6+325738 & 12 58 58.60329 & 32 57 38.1592 & 91.48 & 34\hspace{1.8mm} & 46.0 \# & 47.1 & 17.8 \# 
& 48.7 & AS \\ 

J125900.0+330617 & 12 59 00.0167 & 33 06 17.130 & 277.85 & 144\hspace{1.8mm} & 11.4 \& & 58.7 & $\leq$ 
0.8\hspace{3.8mm} \\ 

J125954.0+335653 & 12 59 54.01264 & 33 56 53.3788 & 478.02 & 146\hspace{1.8mm} & 74.1 @ & 153.3 & 7.9 @ 
& 85.1 & AS \\ 

J130056.5+314443 & 13 00 56.5325 & 31 44 43.726 & 82.34 & 31\hspace{1.8mm} & 6.6\hspace{3.8mm} & 14.0 & 
$\leq$ 1.1\hspace{3.8mm} & & S \\ 

J130114.5+313254 & 13 01 14.51914 & 31 32 54.6472 & 46.77 & 21\hspace{1.8mm} & 7.8 \# & 10.0 & 6.6 \# & 
7.2 & A \\ 

J130121.4+340030 & 13 01 21.42787 & 34 00 30.2497 & 36.94 & 27\hspace{1.8mm} & 30.1 \# & 31.1 & 16.2 \# 
& 29.9 & AS \\ 

J130129.1+333700 & 13 01 29.15475 & 33 37 00.3839 & 75.89 & 52\hspace{1.8mm} & 62.1 \# & 62.8 & 123.2 \# 
& 127.0 & AS \\ 

J130137.5+323423 & 13 01 37.58463 & 32 34 23.6414 & 47.50 & 21\hspace{1.8mm} & 10.9 \# & 12.8 & 3.4 \# & 
13.1 & S \\ 

J130150.4+313742 & 13 01 50.44960 & 31 37 42.9140 & 215.65 & 60\hspace{1.8mm} & 12.1 \& & 37.2 & 
1.1\hspace{3.8mm} &  $-$ \\ 

J130200.7+340232 & 13 02 00.7554 & 34 02 32.164 & 65.45 & 28\hspace{1.8mm} & 3.0\hspace{3.8mm} & 8.9 & 
$\leq$ 0.6\hspace{3.8mm} \\ 

J130230.8+305109 & 13 02 30.8975 & 30 51 09.299 & 45.37 & --\hspace{1.8mm} & 2.8\hspace{3.8mm} & 6.5 & 
$\leq$ 0.8\hspace{3.8mm} \\ 

J130310.2+333406 & 13 03 10.27119 & 33 34 06.7751 & 100.45 & 44\hspace{1.8mm} & 5.8 \# & 8.4 & 4.5 \# & 
4.8 & AS \\ 

J130311.8+320739 & 13 03 11.87280 & 32 07 39.3167 & 30.35 & 32\hspace{1.8mm} & 13.2 \# & 14.4 & 12.0 \# 
& 16.0 & AS \\ 

J130411.2+310222 & 13 04 11.2967 & 31 02 22.631 & 34.26 & --\hspace{1.8mm} & 2.1\hspace{3.8mm} & 4.3 & 
$\leq$ 0.7\hspace{3.8mm} \\ 

J130424.9+334924 & 13 04 24.9873 & 33 49 24.297 & 132.98 & 55\hspace{1.8mm} & 6.7 \& & 33.1 & $\leq$ 
1.0\hspace{3.8mm} \\ 

J130533.1+312327 & 13 05 33.11925 & 31 23 27.3718 & 180.09 & 67\hspace{1.8mm} & 15.9 \& & 35.9 & 
1.5\hspace{3.8mm} & $-$  \\ 

J130620.5+324522 & 13 06 20.5690 & 32 45 22.088 & 52.55 & 20\hspace{1.8mm} & 4.3 \& & 15.2 & $\leq$ 
1.2\hspace{3.8mm} & & AS \\ 

J130650.3+315054 & 13 06 50.3492 & 31 50 54.812 & 40.96 & --\hspace{1.8mm} & 4.7\hspace{3.8mm} & 8.4 & 
$\leq$ 1.1\hspace{3.8mm} & & S \\ 
  \hline
  \end{tabular}
\\
Notes: Col.~1 -- source name, derived from the best available radio coordinates; 
Col.~2 -- J2000 right ascension (h~m~s); Col.~3 -- J2000 declination 
($\degr$~$\arcmin$~$\arcsec$) (the accuracy of the improved coordinates 
for the MERLIN-only and VLBI detections is different; see the text for the 
details); Col.~4 -- FIRST integrated flux density at 
1.4~GHz (mJy); Col.~5 -- GB6 total flux density at 5~GHz (mJy) (Gregory et al. \cite{gb6}),
$^{a, b}$ mark double sources unresolved in GB6 (see also Table~\ref{M-non});
Col.~6 -- MERLIN peak brightness at 5~GHz
(mJy/beam); the MERLIN image is displayed in Fig.~\ref{vplot1} (\#),
Fig.~\ref{vplot3} (@) or Fig.~\ref{mplot1} (\&); Col.~7 -- MERLIN
integrated flux density at 5~GHz (mJy); Col.~8 -- VLBI peak brightness at
5~GHz (mJy/beam), an upper limit is given for VLBI non-detections; the
VLBI image is displayed in Fig.~\ref{vplot1} (\#) or Fig.~\ref{vplot3}
(@); Col.~9 -- VLBI integrated flux density at 5~GHz (mJy), marginal VLBI
detections are indicated with a ``$-$'' sign. Col.~10 -- optical 
identifications of the sources: SDSS DR4 (S) and 
APM (A), McMahon et al. (\cite{mcmahon}).
\end{table*}

\subsection{MERLIN filter observations and data reduction}

After selecting the reference source and the targets, we imaged the sample
at an intermediate angular resolution with MERLIN.
The primary goal of the MERLIN observations was to optimise the subsequent
VLBI observations: to identify the objects that may be suitable for the
highest angular resolution ($\sim 1$~mas) imaging, and to improve the
astrometric source positions. The NGP field was observed with MERLIN at
5~GHz with 14 MHz bandwidth in dual circular polarisation on 24--28 March
2001. The total observing time was 29.8~hours. Each target source was
observed 6 times in 4.5-min scans between 2-min intervals spent on the
phase-reference calibrator, including antenna slewing.
The resulting image angular resolution was $\sim 50$~mas.
The data were
calibrated with the NRAO AIPS package (e.g. Diamond \cite{aips}) and 
imaged with AIPS and the Caltech Difmap package (Shepherd et al. 
\cite{difmap}). A large ($5\arcmin \times 5\arcmin$) field of view was
investigated in each case to look for distant source components. In the 
NGP field, 
37 out of the total 47 sources have been detected with MERLIN and  
proposed later for VLBI observations (Table~\ref{sample_ordered}).
For all of these objects, the MERLIN image peak brightness exceeded 
$\sim2$~mJy/beam ($3\sigma$). The ten sources from the originally defined 
NGP 
sample that were not detected with MERLIN are listed in Table~\ref{M-non}, 
along with their FIRST integrated flux density at 1.4~GHz and the upper 
limit of their MERLIN peak brightness at 5~GHz. These sources appeared 
completely resolved at $\sim 50$~mas angular scale.

\begin{figure}[b]
 \centering
  \includegraphics[width=80mm,bb= 0 0 496 503,clip=]{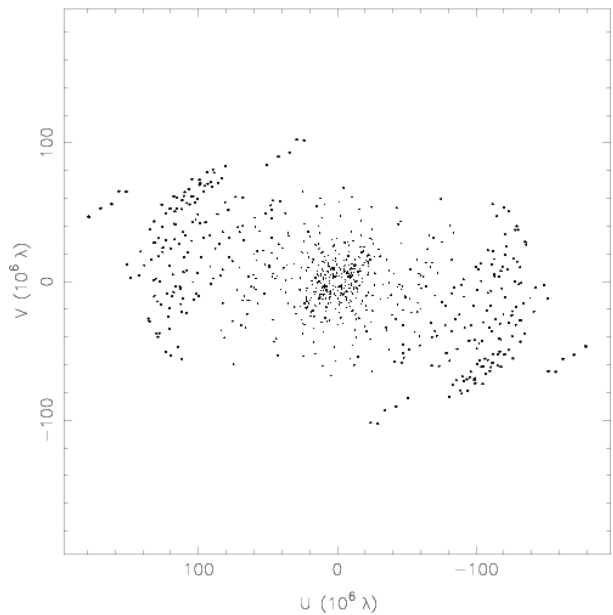}
  \caption{The $(u,v)$-coverage of one of the
program sources (J130311.8+320739) in our 5-GHz global VLBI observations.}
\label{uvpl}
\end{figure}

\subsection{VLBI observations}

The 5-GHz global VLBI observations of the MERLIN-filtered NGP sample
involving 19 antennas of the European VLBI Network (EVN)
and the NRAO Very Long Baseline Array (VLBA)
took place on 30 May 2002. The total observing time was 21 hours, with 7
hours overlap between the two arrays.
The participating radio telescopes
were Effelsberg (Germany), Westerbork (The Netherlands), Jodrell Bank Mk2 
(UK),
Medicina, Noto (Italy), Toru\'{n} (Poland), Onsala (Sweden), Sheshan, Nanshan
(China) from the EVN, and all the ten stations of the VLBA.
The data were recorded with a data rate of 256 Mbit s$^{-1}$, in four 8 
MHz wide  
intermediate frequency channels (IFs) in both left and right circular 
polarisation at all antennas except Sheshan where only left circular
polarisation was available. The total bandwidth was 32~MHz in both
polarisations.

Each programme source was observed in 10 scans of 2.5~min duration.
Taking the slewing times of the antennas into account, it corresponds to 20~min
accumulated observing time per source.
The estimated image thermal noise ($1\sigma$) was $\sim100$~$\mu$Jy/beam.

In each 6-min observing cycle, two programme sources were observed, chosen 
with the minimum angular distance between them. The reference source was 
observed for 1~min per cycle. An example of a ($u,v$)-coverage for a 
program source is shown in Fig.~\ref{uvpl}. The typical restoring beam 
using natural weighting is $\sim1$~mas~$\times$~3~mas in a position angle 
(PA) of $-16\degr$.

\begin{table*}
 \caption{Data of the 10 DEVOS NGP sources not detected with MERLIN.}
 \label{M-non}
 \begin{tabular}{@{}lcccc}
  \hline
  Source name & FIRST flux    & GB6 flux      & MERLIN peak & Opt. ID\\
              & density (mJy) & density (mJy) &  (mJy/beam)  \\
  \hline
J125102+315634 &  44.35 & 33\hspace{2mm} & $\leq$1.9 & A \\ 
J125543+324115 &  33.99 & 34\hspace{2mm} & $\leq$2.0 \\ 
J125546+304605 & 138.37 & $81^{a}$ & $\leq$2.2 \\ 
J125746+305952 &  38.04 & 20\hspace{2mm} & $\leq$2.3 & A \\ 
J125821+313130 &  46.53 & --\hspace{2mm} & $\leq$1.7 \\ 
J125938+313047 &  53.70 & 24\hspace{2mm} & $\leq$1.7 \\ 
J130203+320635 &  74.02 & --\hspace{2mm} & $\leq$1.9 \\ 
J130334+305817 &  33.99 & --\hspace{2mm} & $\leq$1.7 \\ 
J130557+333016 &  39.58 & $33^{c}$ & $\leq$1.6 & A \\ 
J130557+333030 &  42.35 & $33^{c}$ & $\leq$1.8 \\ 
  \hline
  \end{tabular}
\\
Notes: Col.~1 -- source name; Col.~2 -- FIRST integrated flux density at 
1.4~GHz (mJy); Col.~3 -- GB6 total flux density at 5~GHz (mJy) (Gregory et al. \cite{gb6}), 
$^{a, c}$ mark double sources unresolved in GB6 (see also
Table~\ref{sample_ordered});
Col.~4 -- upper limit of MERLIN peak brightness at 5~GHz 
(mJy/beam) corresponding to $3\sigma$ image noise. Col.~5 -- optical 
identifications of the sources: APM (A), McMahon et al. (\cite{mcmahon}).
\end{table*}

\begin{table*}
 \caption{List of the DEVOS NGP sources identified with SDSS imaging objects.}
 \label{opt}
 \begin{tabular}{@{}llcccc}
  \hline
  Source name & SDSS name    & separation ($\arcsec$) & $r$ magnitude & morphology & VLBI-detected\\
  \hline
J125240.2+331058 & SDSS J125240.29+331058.1 & 0.12 & 19.28 & G & $\star$ \\ 
J125354.0+311556 & SDSS J125354.05+311558.9 & 0.28 & 21.37 & G &  \\ 
J125410.8+330912 & SDSS J125410.83+330912   & 0.07 & 22.40 & G & $\star$ \\ 
J125658.4+325053 & SDSS J125658.42+325053.2 & 0.86 & 21.40 & Q & $\star$ \\ 
J125728.6+342428 & SDSS J125728.64+342429.1 & 0.20 & 22.53 & G &  \\ 
J125755.7+313915 & SDSS J125755.77+313915.2 & 0.59 & 23.10 & G & $\star$ \\ 
J125858.6+325738 & SDSS J125858.6+325738.2  & 0.09 & 19.41 & Q & $\star$ \\ 
J125954.0+335653 & SDSS J125954.01+335653.3 & 0.07 & 20.74 & G & $\star$ \\ 
J130056.5+314443 & SDSS J130056.49+314442.7 & 1.08 & 22.54 & G &  \\ 
J130121.4+340030 & SDSS J130121.42+340030.1 & 0.06 & 19.32 & Q & $\star$ \\ 
J130129.1+333700 & SDSS J130129.15+333700.3 & 0.03 & 19.68 & Q & $\star$ \\ 
J130137.5+323423 & SDSS J130137.6+323423.2  & 0.43 & 23.20 & G & $\star$ \\ 
J130310.2+333406 & SDSS J130310.26+333406.7 & 0.03 & 19.96 & Q & $\star$ \\ 
J130311.8+320739 & SDSS J130311.87+320739.3 & 0.05 & 20.41 & Q & $\star$ \\ 
J130620.5+324522 & SDSS J130620.56+324522   & 0.06 & 18.66 & G &  \\ 
J130650.3+315054 & SDSS J130650.33+315054.4 & 0.42 & 22.35 & G &  \\ 
  \hline
  \end{tabular}
\\
Notes: Col.~1 -- source name; Col.~2 -- SDSS source name; 
Col.~3 -- angular distance between the best available radio and the SDSS optical positions ($\arcsec$);
Col.~4 -- SDSS $r$ magnitude;
Col.~5 -- optical morphology: extended (galaxy, G) or point-like (quasi-stellar object, Q);
Col.~6 -- VLBI-detected sources ($\star$).
\end{table*}

\section{VLBI data processing and imaging}
\label{sect3}

\begin{figure}
\centerline{
  \includegraphics[width=80mm,bb= 32 156 583 672,clip= ]{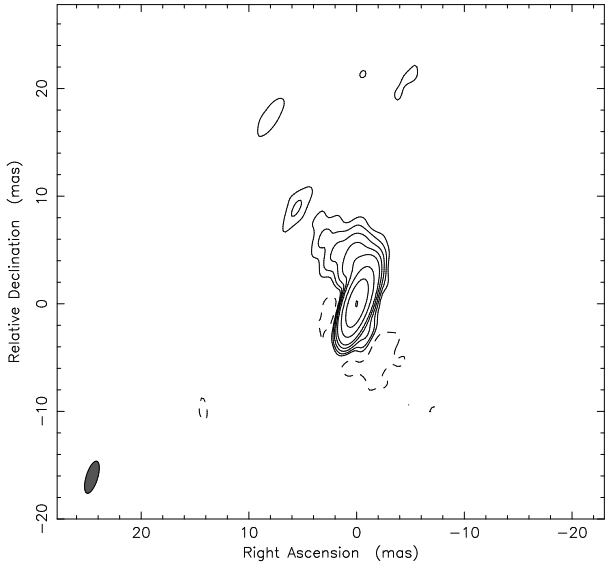}}
  \caption{Naturally weighted 5-GHz VLBI image of the reference source 
J1257+3229. The lowest contours are drawn at $\pm0.61$~mJy/beam. The
positive contour levels increase by a factor of 2. The peak brightness is
645~mJy/beam. The restoring beam is $3.1$~mas~$\times$~$1.1$~mas at
PA=$-17\degr$. The coordinates are relative to the brightness peak. The
Gaussian restoring beam is indicated with an ellipse in the lower-left corner.}
\label{calib}
\end{figure}

The correlation took place at the EVN MkIV Data Processor at the Joint 
Institute for VLBI in Europe (JIVE) in Dwingeloo (The Netherlands). The 
NRAO AIPS package was used for the initial data calibration. The reference 
source J1257+3229 was imaged first (Fig.~\ref{calib}).
The residual amplitude and phase corrections resulted from the hybrid mapping
of J1257+3229 were applied to the target sources. No amplitude 
self-calibration was attempted after this point except in the cases of 
the two brightest target sources (J125240.2+331058 and J130129.1+333700).
The original field of view of VLBI images limited by time-smearing was $4\farcs6$.

\setcounter{figure}{5}
\begin{figure}[t]
\centering
\includegraphics[width=8.0cm,bb= 0 20 270 220,clip= ]{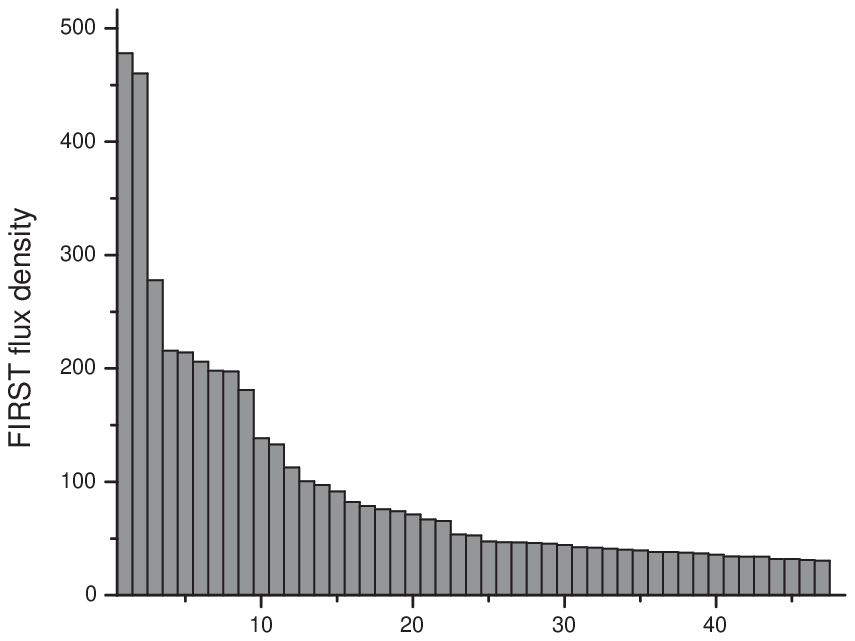}
\includegraphics[width=8.0cm,bb= 0 20 270 230,clip= ]{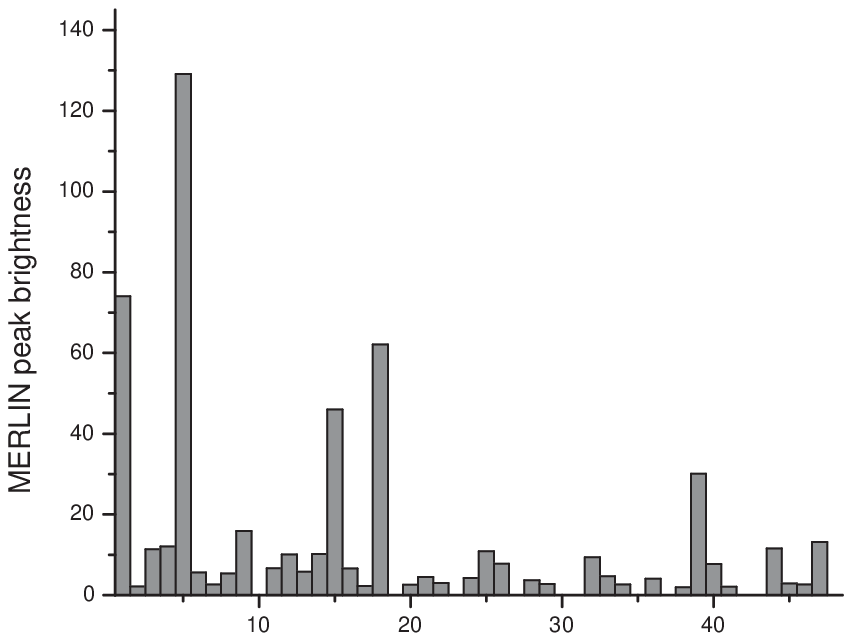}
\includegraphics[width=8.0cm,bb= 0 20 283 230,clip= ]{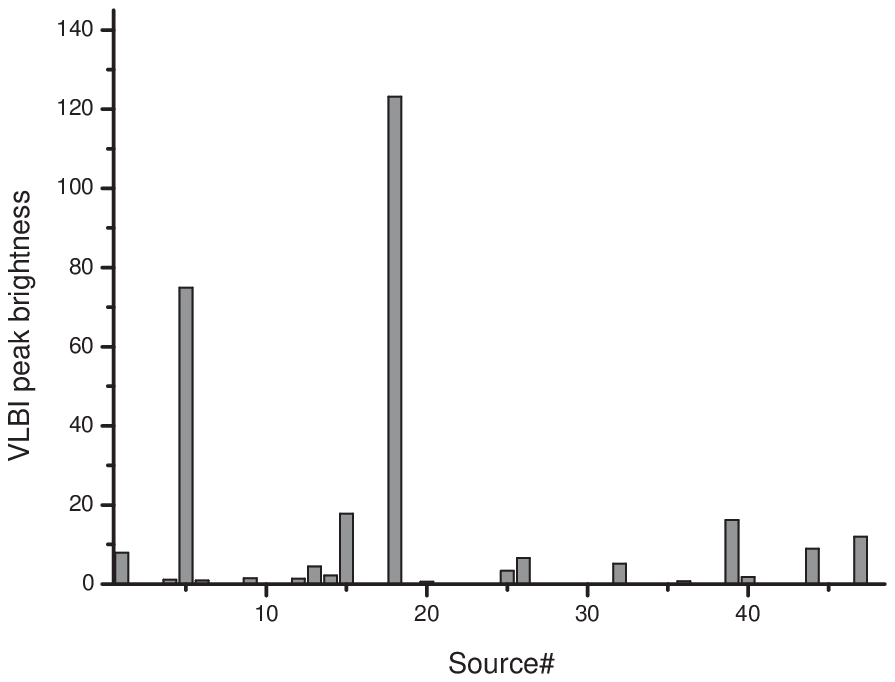}
\caption{The FIRST 1.4-GHz integrated flux densities (mJy) of the 47 
DEVOS NGP sample sources in decreasing order ({\it top}), and the 
corresponding 5-GHz MERLIN ({\it middle}) and VLBI ({\it bottom}) peak 
brightnesses (mJy/beam) of each source. Note that one of the sources 
($\#18$: J130129.1+333700) showed significant flux density variability, 
see Sect.~\ref{sect4_notes}.}
\label{dist1}
\end{figure}

\begin{figure}[t]
  \centering
 \includegraphics[width=80mm]{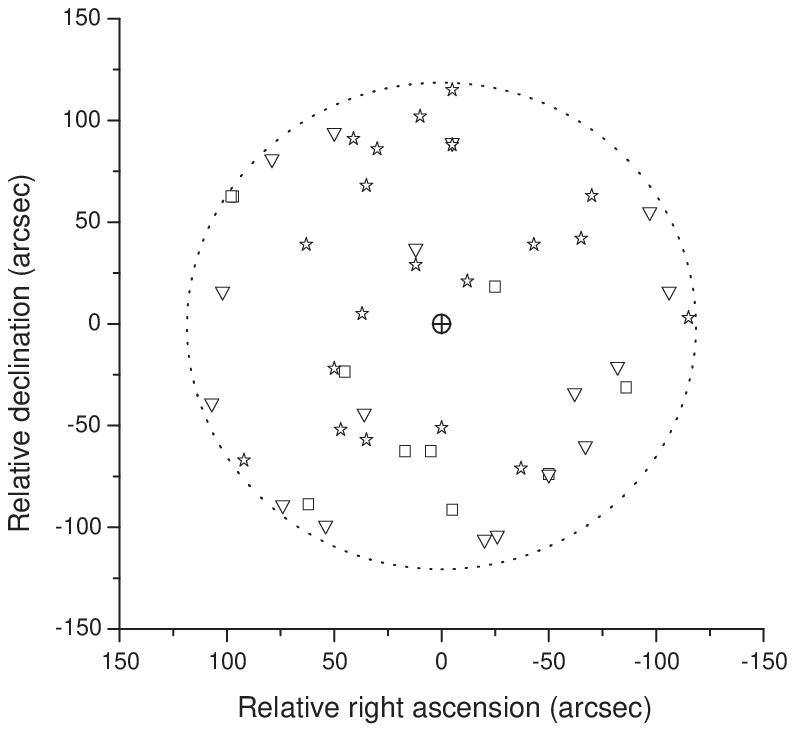}
  \caption{The positions of the program sources with respect to the
phase-reference calibrator source J1257+3229. The dashed circle shows 
the $2{\degr}$ selection radius. The VLBI-detected sources (19) are marked
with stars, the MERLIN-only detections (18) with triangles,
and the MERLIN non-detections (10) with squares.}
  \label{throw}
\end{figure}

\begin{figure}
  \centering
   \includegraphics[width=8cm]{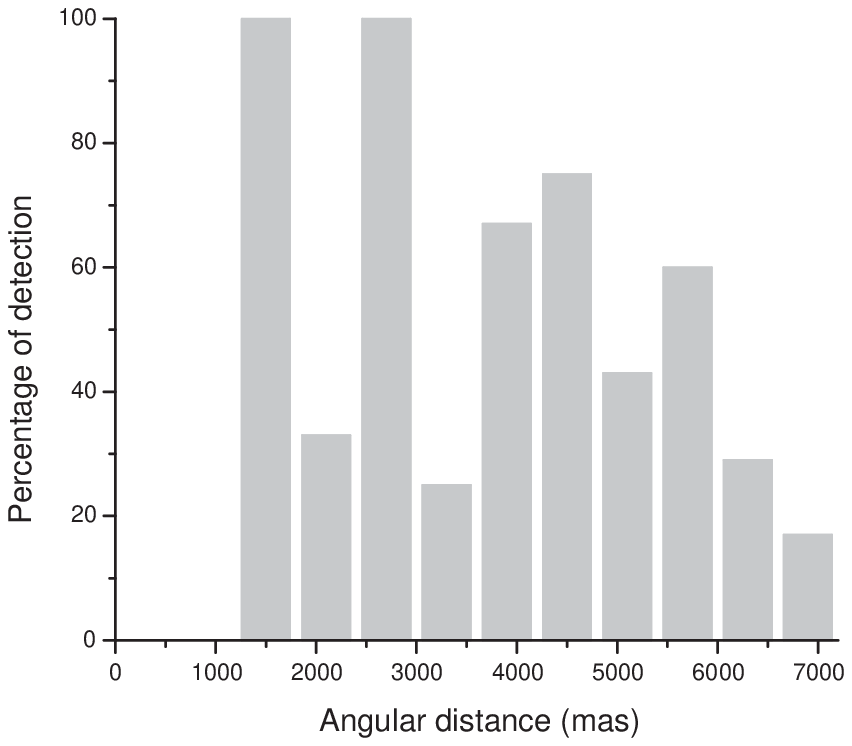}
  \caption{The percentage of the detected with respect to the observed
sources with VLBI as a function of the
angular separation from the phase-reference calibrator source.
The sources are grouped into equally wide 600-mas concentric rings.}
  \label{ratio-throw}
\setcounter{figure}{2}
\end{figure}

Initially, all programme sources were imaged and investigated 
individually.
Difmap and AIPS packages were used for imaging. For the sources brighter 
than $\sim15$ mJy/beam, phase self-calibration was performed.
For the rest of the sources,
the images are the results of a CLEANing process only.
For phase self-calibration, we used a simple initial
brightness distribution model, the CLEAN model of the phase-reference 
image for each source, with a 3-min solution interval. 
The information on the target source positions relative to the reference 
source is preserved this way.
We carried out tests on the reliability of
the imaging by means of comparing the results obtained with both packages.
In Difmap, we do not have full control over the self-calibration process 
in a sense that one cannot set a signal-to-noise ratio (SNR) threshold for 
the solutions to be accepted or dismissed. In AIPS, we were able to set 
SNR$\geq5$ and apply phase self-calibration only if most of the solutions 
were good. Our experience showed that the images made using both packages
this way were equivalent. 

With the ultimate aim of observing and imaging thousands of radio sources 
in the full DEVOS, an automated (fast and objective) procedure to
prepare the VLBI images is needed. Relatively simple scripts have been 
written and tested for this purpose using Difmap. We then compared the 
images prepared ``manually'' and reproduced in this paper with those 
obtained by the automated procedure.
It turned out that the same sources were detected, with qualitatively 
identical mas-scale structures. The peak brightness and image noise values 
were also consistent. The differences in peak brightnesses were typically 
less than $\pm10$\%.

\section{Results and discussion}
\label{sect4}

The sample of 37 MERLIN-detected sources observed with VLBI is given in 
Table~\ref{sample_ordered}. Among them, a total of 19 sources were 
detected and imaged with VLBI. Natural weighting of the visibility data 
was used to maximise the sensitivity. The MERLIN and 
VLBI images of 12 sources are shown in Fig.~\ref{vplot1}. The images of 
the complex structure of one more source, J125954.0+335653 are shown in 
Fig.~\ref{vplot3}. The detailed image parameters are given in 
Table~\ref{VLBI-maps}. Additionally, there were 6 marginal VLBI detections
located at the MERLIN position within the MERLIN source structure. These 
cases are indicated with ``$-$'' in Col.~9 of 
Table~\ref{sample_ordered}.
For these sources, only the dirty images were investigated but not
reproduced in this paper since they all show simple featureless
low-brightness structure. For the rest of the VLBI-observed sources, the
upper limits of the peak brightness corresponding to the $3\sigma$ image
noise are shown in Table~\ref{sample_ordered}.

The majority (17 out of 24) of the sources not detected or marginally
detected with VLBI appeared compact and unresolved with MERLIN. The
images of these sources are not shown here but the MERLIN peak brightness
values are reported in Table~\ref{sample_ordered}. However, 7 of the 24
sources exhibit significantly resolved structure at the MERLIN angular
scales (Fig.~\ref{mplot1}).

In Tables~\ref{sample_ordered} and \ref{M-non}, we also list the 5-GHz total flux densities taken 
from the GB6 catalogue (Gregory et al. \cite{gb6}). 
The radio spectral properties of the sources can be inferred from comparing these data with 
the FIRST flux densities. However, a large fraction of the sources in our sample is not found in GB6: 
their 5-GHz total flux densities are lower than the 18-mJy detection threshold there. Since our survey 
is more sensitive, five of these objects have been detected even with VLBI, with integrated flux densities 
up to 13.6~mJy. The GB6 data would have limited potential in constraining our initial sample.

We looked for close companions of our sources selected as unresolved in FIRST. It turned out that many of the 
objects that we resolved, 8 among the 10 sources not detected with MERLIN (Table~\ref{M-non}), have a companion 
within $\sim 1 \arcmin$ in the FIRST catalogue. These pairs of sources are most likely physically associated, 
although chance coincidences in about half of the cases are also possible at around $1\arcmin$ angular separation 
(cf. McMahon et al. \cite{mcmahon}). There are 14 such ``double'' sources in the DEVOS NGP sample, and only one of 
them was -- marginally -- detected with VLBI (J125734.2+335801). However, a compact source in a FIRST ``triple'' 
(J125240.2+331058) is a strong VLBI detection. We shall investigate whether the initial sample selection could be 
usefully constrained by excluding the components of ``double'' FIRST sources in the future.

The distribution of MERLIN and 
VLBI peak brightnesses are shown in Fig.~\ref{dist1}. Based on their peak brightness, the 19 sources 
detected with VLBI can be grouped as follows:

\begin{itemize}
 \item 2 bright objects ($S_{\rm{peak}}\sim$75 and 123~mJy/beam)
 \item 3 objects with $S_{\rm{peak}}\sim10-20$~mJy/beam
 \item 8 objects with $S_{\rm{peak}}\sim1-8$~mJy/beam
 \item 6 objects marginally detected with $\sim~1$~mJy/beam 
peak brightness ($\sim4\sigma$, with possibly extended structures resolved 
with VLBI) and co-located within the MERLIN source structure.
\end{itemize}

In certain cases the image noise achieved is higher than the thermal
noise expected (see Col. 4 of Table~\ref{VLBI-maps}). 
The dynamic range may be limited by residual phase-referencing
errors introduced by the switching technique.
Direct fringe-fitting was attempted for four target sources brighter than 
$\sim 15$~mJy/beam in order to estimate the coherence loss in the 
phase-referencing process by comparing the images obtained in both ways. 
The differences of peak brightnesses ranged from 5 to 25\%. 
Therefore the $3\sigma$ threshold for the VLBI non-detections was
multiplied by a factor of 1.25 in Col.~8 of Table~\ref{sample_ordered}
to account for this effect.

We note that we were able to image the faintest source (J130311.8+320739)
in our initial FIRST sample. The object shows a compact radio structure
with 10-mJy level flux density at 5~GHz.
The detections are apparently not affected by the angular distance of the
target sources from the phase-reference calibrator (Fig.~\ref{throw} and 
\ref{ratio-throw}). For
example, the object J124833.1+323207 is located at the very edge of the
$2\degr$-radius circle around the reference source. It was detected with
a peak brightness of 5.2~mJy/beam. The VLBI detection rate is not
significantly different for small and large target--reference separations
(Fig.~\ref{ratio-throw}). We also note that for 2 VLBI-detected sources 
(J125755.7+313915 and J130114.5+313254) the deconvolved FIRST angular sizes 
are $\sim5\arcsec$ (4.91$\arcsec$ and 4.85$\arcsec$, respectively). 

The improved VLBI coordinates of the brightness peak locations for the 
sources
detected with MERLIN and VLBI are given in Table~\ref{sample_ordered}.
The source names are derived from these coordinates according to the IAU 
convention.
The formal positional accuracy is proportional to the angular resolution 
of the array and scales inversely with the signal-to-noise ratio,
i.e. the source strength (e.g. Lestrade et al. \cite{lestrade}). For the 
MERLIN-only detections, the accuracy estimated is 9~mas ($3\sigma$, for
the weakest sources) or better in both $\alpha$ and $\delta$.
For sources clearly detected with VLBI (those with integrated VLBI flux 
densities cited in Col.~9 of Table~\ref{sample_ordered}), the 
coordinates relative to those of the phase-reference calibrator source are 
accurate to 0.2~mas in $\alpha$ and 0.4~mas in $\delta$.
These are conservative estimates calculated for weak sources and assuming 
relatively large target--reference source separations. The relative 
astrometric position errors of the target sources in phase-referencing 
observations are known to depend on their angular separation from the 
calibrator. Chatterjee et al. (\cite{chatterjee}) give an estimate of this 
effect by comparing measured and modelled pulsar positions. Their results 
indicate that the astrometric errors are within 0.5~mas at 5~GHz frequency 
even at $2\degr$ separation, the limit we used in our experiment. Our 
estimates are consistent with the results of Chatterjee et al. 
(\cite{chatterjee}). Taking the accuracy of the phase-reference calibrator 
coordinates into account, the absolute ICRF positions of the VLBI-detected 
DEVOS sources are accurate to at least 0.6~mas and 0.7~mas in $\alpha$ and 
$\delta$, respectively. For marginal VLBI detections, these absolute 
accuracies are $\sim1$~mas.

Although the number of observed and detected sources in our pilot
experiment is rather small, it is still possible to get an idea for the
VLBI detection rate of FIRST-based source samples selected using similar
critera to those employed here. Taking the results of Garrington et al.
(\cite{gg99}) and Wrobel et al. (\cite{wrobel}) into account as well, a
VLBI detection rate of up to 40\% is expected.

Most recently, Porcas et al. (\cite{porcas}) conducted a survey of $\sim 1000$
FIRST radio sources at 1.4~GHz with a single highly sensitive VLBI baseline
between Arecibo and Effelsberg.
They found that about one third of all sources
had detectable compact structure at the mJy level. This ratio holds for the
weakest ($\sim 1$ mJy) sources as well, and is essentially independent from
the source total flux density.

A deep wide-field VLBI survey was conducted by Garrett et al.
(\cite{garrett04b}) at 1.4~GHz with the VLBA and the 100-m Green Bank Telescope.
Sixty-one mJy and sub-mJy-level radio sources previously imaged with the
Westerbork Synthesis Radio Telescope (WSRT) at arcsecond scales
were observed simultaneously. A total of 9 sources were detected with
VLBI, with $1\sigma$ image noise ranging from 9 to 55~$\mu$Jy/beam. 
The detection rate is 29\% for the mJy WSRT sources, giving a lower limit 
for the fraction of the objects that are powered by accretion process onto 
massive black holes. This detection rate is consistent with our result.

\subsection{Notes on individual sources}
\label{sect4_notes}

The VLBI images of the sample sources reveal a large variety of radio 
morphologies, ranging from those typical for compact AGNs through 
``classical'' core--jet structures to highly resolved sources. 
Cross-identification of the DEVOS pilot sample objects using
imaging data from the SDSS revealed optical counterparts 
of 16 sources of the total 47 (34\%). Details are given in 
Tables~\ref{sample_ordered} and \ref{opt}. All of the optically identified 
radio sources were detected with MERLIN and 11 also with VLBI 
(including 2 marginal detections). All the 6 sources that appear point-like
in the SDSS images were detected with VLBI and thus are among the most 
compact objects in the radio as well. These quasi-stellar objects would be 
classified by spectroscopic data conclusively.
Eight optical counterparts within $5\arcsec$ of our 19 
VLBI-detected sources were found in the catalogue of Automatic Plate 
Measuring (APM) identifications using the Palomar Observatory Sky Survey 
(POSS) first-epoch red and/or blue plates (McMahon et al. 
\cite{mcmahon}, see Col.~10 of Table~\ref{sample_ordered}). 
Moreover, 3 out of our 18 MERLIN-only detections (also Col.~10 of 
Table~\ref{sample_ordered}) and 3 out of 10 MERLIN 
non-detections also have APM counterparts (Col.~5 of Table~\ref{M-non}).
Here we briefly comment on sources that have interesting observational
properties or additional information available from the literature.

{\bf J125240.2+331058} The FIRST image shows a three-component radio structure.
We were able to image with VLBI the central compact object that lies between 
two diffuse low-brightness lobes $\sim 1\arcmin$ apart from each other.

{\bf J125858.6+325738} Based on the structural pattern, this source 
might be a compact symmetric object (CSO, cf. Taylor et al. \cite{taylor}).

{\bf J125954.0+335653} This galaxy is the strongest source at 1.4~GHz in
the DEVOS NGP sample. A redshift of $z=0.63$ was measured by Allington-Smith
et al. (\cite{allington}). It was resolved by MERLIN into a double-lobe radio structure.
We could detect both
components separated by $\sim240$~mas with VLBI, although heavily resolved
(Fig.~\ref{vplot3}). This corresponds to a projected linear extent of
$\sim1.6$~kpc, assuming a flat cosmological model with $\Omega_{m}=0.3$,
$\Omega_{\Lambda}=0.7$ and $H_{\rm{0}}=70$~km~s$^{-1}$~Mpc$^{-1}$.
Based on the steep spectrum (radio spectral index $-0.6$)
and the small size,
the object can be identified as a CSS source. Its 5-GHz rest-frame radio luminosity is
$4.2\times10^{26}$~W~Hz$^{-1}$.

{\bf J130121.4+340030} The VLBI image of this quasar (also known as 
5C~12.165) at a redshift of $z=1.970$ (Barkhouse \& Hall 
\cite{barkhouse}) shows a typical one-sided core--jet structure.

{\bf J130129.1+333700} The integrated 5-GHz flux density of this source
measured in our MERLIN experiment was 62.8~mJy in March 2001. About a 
year later (in May 2002), a flux density of 127.0~mJy was determined with 
VLBI. This indicates significant flux density variability, consistent 
with the compact radio structure. Note that --~according to what is 
expected in the absence of variability~-- none of the other sources 
showed VLBI peak brightness higher than the MERLIN value 
(Fig.~\ref{dist1}).

\section{Summary and future outlook}
\label{sect5}

The primary goal of the DEVOS pilot project was to verify and adjust sample
selection criteria, observing strategies and data reduction procedures before
the full survey has started. Here we presented MERLIN and global VLBI
observations of a sample of 47 radio sources selected from the FIRST 
survey with $S_{1.4} > 30$~mJy and $\theta < 5\arcsec$, within $2\degr$ 
separation from the phase-reference calibrator source J1257+3229. We 
detected 37 sources at 5~GHz with MERLIN observations at a peak brightness 
of at least $\sim2$~mJy/beam, filtering out extended radio structures not
detected with MERLIN.
Subsequent 5-GHz observations with the global VLBI network revealed that 
19 of the sources are stronger than $\sim 1$~mJy at 
an angular resolution of $\sim 1$~mas. The MERLIN images of 20 and
VLBI images of 13 sources are presented. We were able to identify 34\% of 
the pilot source sample with SDSS-detected sources. All of these sources were detected 
with MERLIN and 11 with VLBI. All sources with unresolved optical structure were
detected with VLBI.

With the typically $0.3-0.7$~mJy/beam ($3\sigma$) VLBI image noise achieved,
we could determine the mas-scale brightness distribution of sources with
rest-frame brightness temperatures of at least $\sim5 \times 10^6$~K. 
Hence the objects detected are likely to be powered by AGN rather than 
starburst 
activity (e.g. Condon \cite{condon}). In DEVOS, we are probing the same 
AGN population, but one or two orders of magnitude fainter than 
traditionally surveyed with imaging VLBI.
The results of this pilot study can
already be valuable in their own right since there are a couple of
individual sources that may be worth studying further (e.g. 
J125858.6+325738 and J130129.1+333700).

With the current observational setup in VLBI, the total time spent on the 
sources (20~min) is sufficient to detect and image the compact objects 
with a peak 
brightness of at least $\sim$1~mJy/beam. The phase-referencing technique 
in the ``nodding style'' was successful: the 6-min long observing cycles 
involving scans on two different target sources between the calibrator 
scans proved adequate at 5~GHz.

Tools are available for efficient data reduction in a highly automated 
way for the full survey. In this pilot project, we used relatively simple 
Difmap scripts which were tested with our MERLIN and VLBI data.
The results were carefully compared with manually prepared AIPS and 
Difmap images. We also aim at developing an AIPS pipeline procedure
for calibration and data reduction in the future such as e.g.
the EVN pipeline (Reynolds et al. \cite{reynolds}).

Potential phase-reference calibrator sources are well distributed in the
sky, thus practically any celestial area north of $-30\degr$ 
declination can be covered in DEVOS. Emphasis
should be made on the areas subject to deep optical surveys in order to
supplement the data base with optical identifications and redshift data.
Due to the nature of the phase-referencing observations, high-quality
images of the calibrator sources are obtained, which could be considered
as valuable by-products on otherwise scientifically interesting bright
radio-loud AGNs.

We found that the sources that appear associated with a nearby object 
in the FIRST catalogue have a relatively low chance for being detected at 
sub-arcsecond angular scales.
In our MERLIN observations, there were 11 objects with 5-GHz peak
brightness between 2 and 4~mJy/beam. Only one of them was marginally
detected later with VLBI (J125734.2+335801).
Therefore if a higher threshold of 4~mJy/beam is
set as a criterion for inclusion in the sub-sample observed with VLBI
with this setup, 18 of the 26 objects would have been detected, giving a
considerably higher success rate. Since the DEVOS is a very demanding project in
terms of network resources, the high detection rate is of special importance.

If more stringent filtering criteria are applied, the number of sources
in a field around a given phase-reference calibrator could possibly be
increased by allowing somewhat larger (e.g. $2\fdg5$) maximum
target--reference angular separation. This would allow us to image about
the same number of sources per VLBI experiment. Another possibility to
enlarge the initial FIRST sample is to set a lower limit on the integrated
1.4-GHz flux density, e.g. 20 mJy.

With a detection rate of $\sim40$\%, VLBI imaging of 10\,000 
objects implies 25\,000 sources in the parent FIRST-based sample to 
be observed with MERLIN. The full SDSS-FIRST spectroscopic quasar sample 
will contain $\sim15000$ objects (Ivezi\'c et al. \cite{ivezic}). 
Our experience shows that a significant fraction of these is expected to 
have compact radio structure and could be detected with VLBI.
With the technical capabilities available
for the pilot experiment presented here, such a full survey would require 
approximately 600 days of MERLIN and 450 days of VLBI time. 
The latter could be decreased to 300 days if the more stringent value of 
4~mJy/beam is used for the limiting MERLIN peak brigthness.

However, the \mbox{e-MERLIN} development (Garrington et al. 
\cite{garrington04}) planned to be completed by 2007
will lead to a substantial (a factor of 35) increase in the sensitivity 
at 5~GHz, significantly reducing the observing time required for DEVOS.
With real-time eVLBI now in an experimental phase at the EVN 
(Szomoru et al. \cite{szomoru}), it may also be possible in the near future
to perform preliminary source filtering observations using data taken
on the shortest EVN baselines in real time, thus further decreasing the 
demand on VLBI resources.

With the perspective
of further developments in the VLBI technique (disk-based recording,
eVLBI, e.g. van Langevelde et al.
\cite{langevelde} and references therein; Garrett \cite{garrett05}), a survey
like DEVOS would require much less telescope resources than at present.
Conservatively assuming 1~Gbit~s$^{-1}$ data recording rate --~most likely an
underestimate by a factor of a few at the EVN in ten years time~--,
phase-referenced imaging of $\sim10^4$ sources providing the same image
noise as in this pilot study would take about 75 days total observing 
time.

The next logical phase of programmes like DEVOS is to employ the wide-field
technique to survey the $\mu$Jy radio source population at mas angular 
resolution. Deep wide-field VLBI observations have already 
been conducted reaching the $1\sigma$ rms image noise of 9~$\mu$Jy/beam 
(Garrett et al. \cite{garrett04b}). Even though observations of this kind 
are currently very demanding in terms of correlator capacity and data 
analysis 
resources, they open up the perspective of studying the faint active 
galaxy population at the earliest cosmological epochs and help 
understanding their formation and evolution. DEVOS can provide 
calibrator sources for these observations with accurate positions 
necessary to anchor the surveyed fields to the ICRF.

With the advent of the Square Kilometre Array (SKA),
and using this sensitive instrument in a high-resolution configuration
as part of a VLBI network, a much deeper DEVOS-like survey
would become technically feasible (Gurvits \cite{gurvits04}).

After completing this pilot study, we intend to initiate further VLBI and
target-finding filter observations in the framework of DEVOS.
We already observed 42 FIRST-based sources with MERLIN at 5~GHz in a field 
close
to the celestial equator around the phase-reference calibrator quasar
J1549+0237 (Mosoni et al. \cite{equ}). A total of 30
sources have been selected for future VLBI observations.

\begin{acknowledgements}
We thank the anonymous referee for suggestions 
that helped us to improve the paper.
LM acknowledges the help of the JIVE Support Group.
We thank Anita Richards and Peter Thomasson for their assistance with the 
MERLIN data reduction.
This research was supported by the European Commission's IHP Programme 
``Access to Large-Scale Facilities'', under contracts no.
HPRI-CT-1999-00045 and HPRI-CT-2001-00142,
and the Hungarian Scientific Research Fund (OTKA, grant
no.\ T046097). SF acknowledges the Bolyai Research Scholarship received
from the Hungarian Academy of Sciences.
MERLIN is a National Facility operated by the
University of Manchester at Jodrell Bank Observatory on behalf of PPARC. The
European VLBI Network is a joint facility of European, Chinese, South African
and other radio astronomy institutes funded by their national research
councils. The National Radio Astronomy Observatory (NRAO) is a facility of the
NSF operated under cooperative agreement by Associated Universities, Inc.
This research has made use of the FIRST survey and SDSS data base, the NASA's 
Astrophysics Data System, the NASA/IPAC Extragalactic Database (NED) which
is operated by the Jet Propulsion Laboratory, California Institute of
Technology, under contract with the National Aeronautics and Space
Administration.

\end{acknowledgements}

\Online

\begin{figure*}
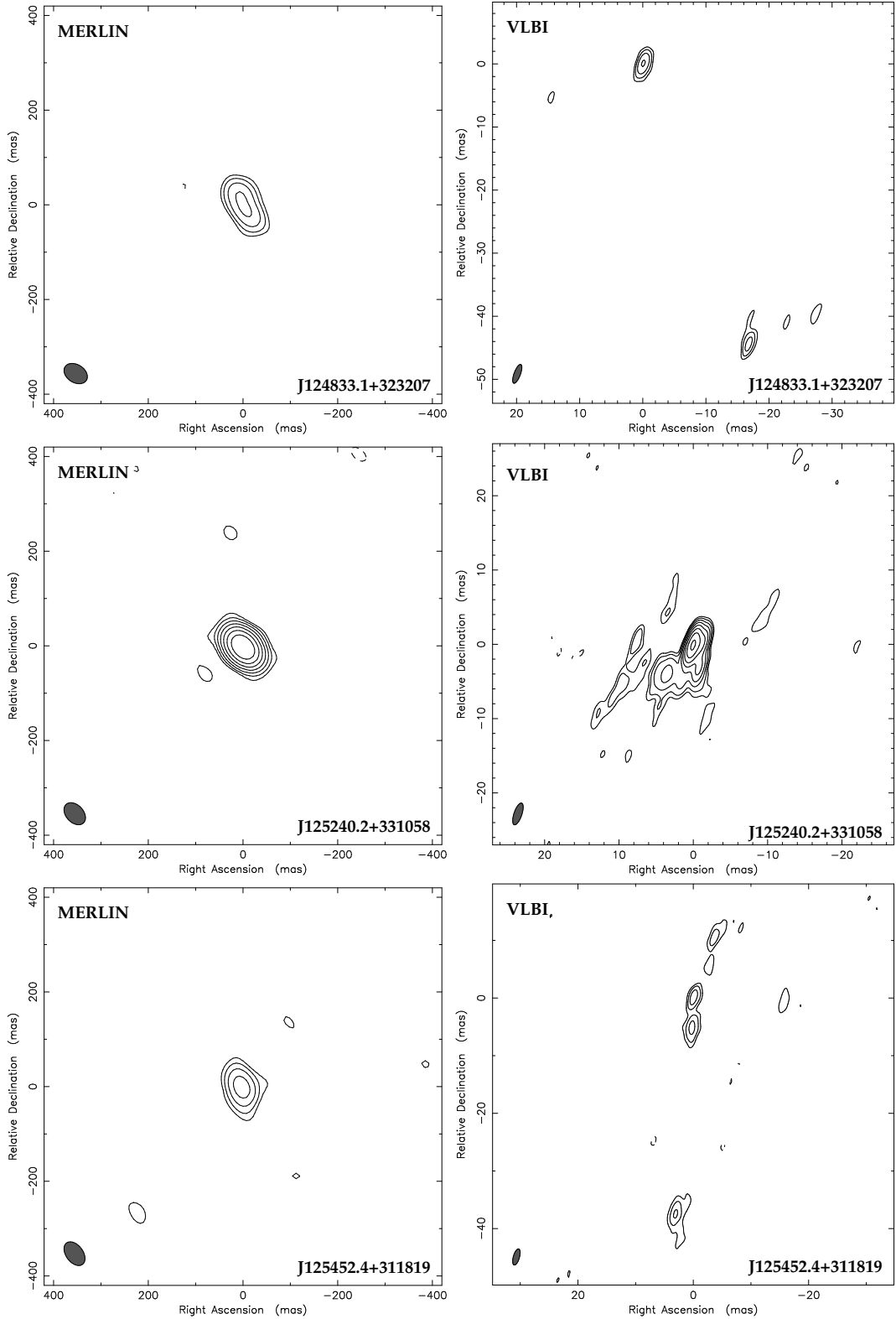

  \centering
    \includegraphics[width=7cm,bb= 32 132 588 680,clip= ]{3473fig3a.ps}
    \includegraphics[width=7cm,bb= 32 132 583 680,clip= ]{3473fig3b.ps}
    \includegraphics[width=7cm,bb= 32 132 588 680,clip= ]{3473fig3c.ps}
    \includegraphics[width=7cm,bb= 32 140 583 688,clip= ]{3473fig3d.ps}
    \includegraphics[width=7cm,bb= 32 132 588 680,clip= ]{3473fig3e.ps}
    \includegraphics[width=7cm,bb= 32 132 583 680,clip= ]{3473fig3f.ps}
  \caption{Naturally weighted 5-GHz MERLIN ({\it left}) and VLBI ({\it 
right}) images of 12 programme sources clearly detected with VLBI. The
image parameters (peak brightness, lowest contour level corresponding to $3\sigma$ image noise,
Gaussian restoring beam size and orientation) are listed and explained in 
Table~\ref{VLBI-maps}. 
The positive contour levels increase by a factor of 2. The coordinates are 
relative to the VLBI brightness peak for which the absolute coordinates 
are given in Table~\ref{sample_ordered}. The restoring beam is indicated 
with an ellipse in the lower-left corner.}
  \label{vplot1}
\end{figure*}

\addtocounter{figure}{-1}
\begin{figure*}
  \centering
    \includegraphics[width=7cm,bb= 32 132 588 680,clip= ]{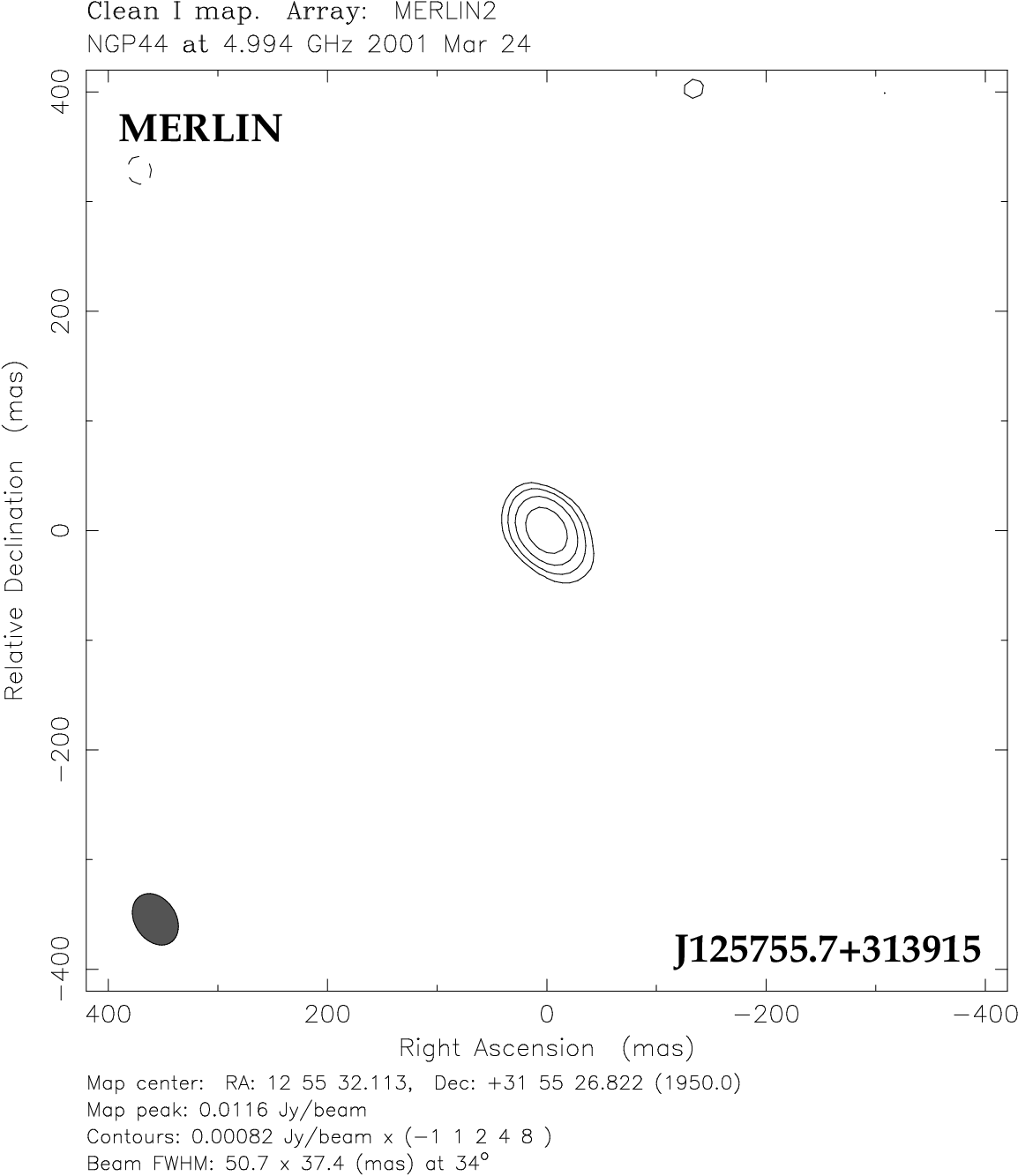}
    \includegraphics[width=7cm,bb= 32 132 583 680,clip= ]{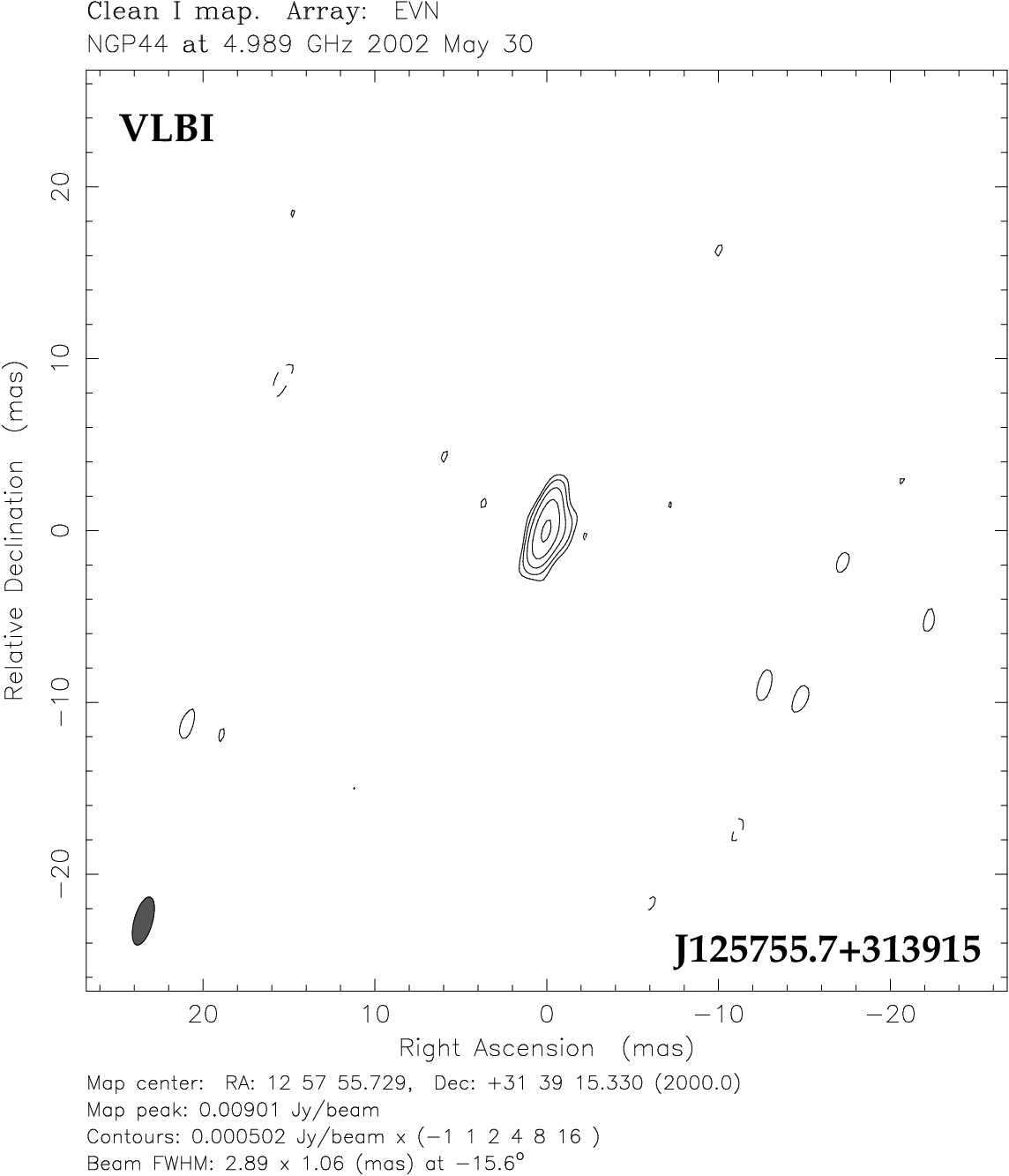}
    \includegraphics[width=7cm,bb= 32 132 588 680,clip= ]{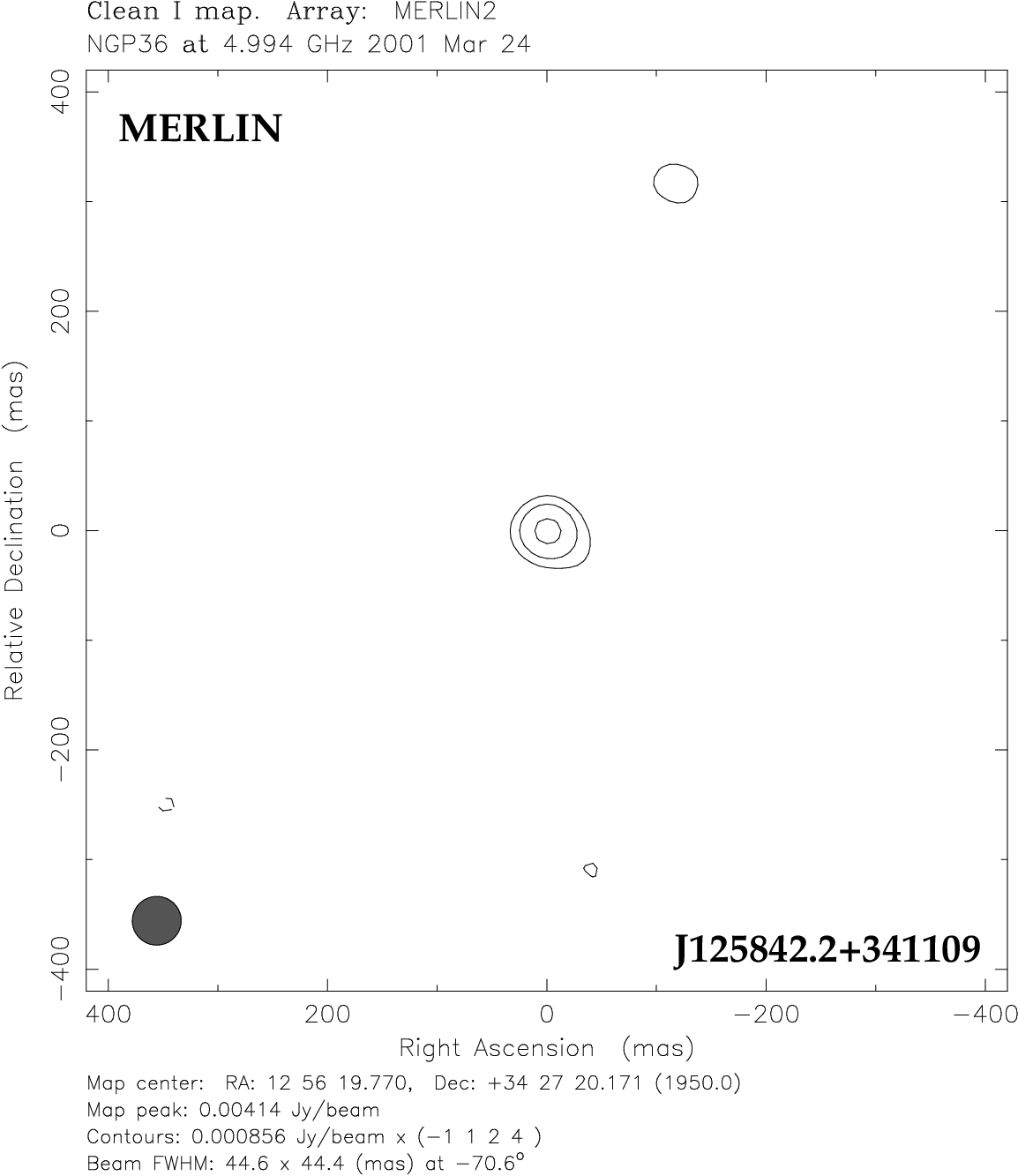}
    \includegraphics[width=7cm,bb= 32 132 583 680,clip= ]{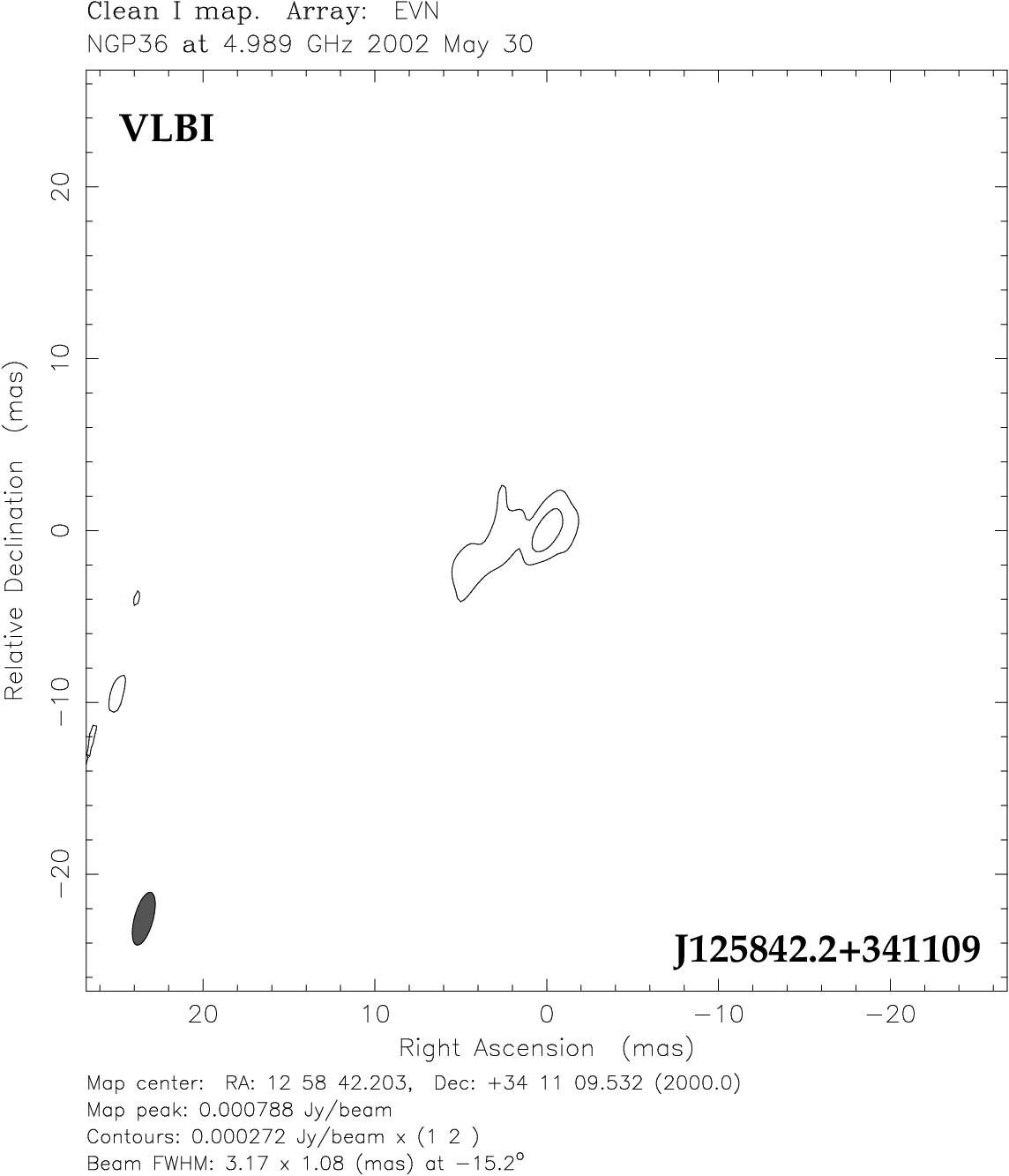}
    \includegraphics[width=7cm,bb= 32 132 588 680,clip= ]{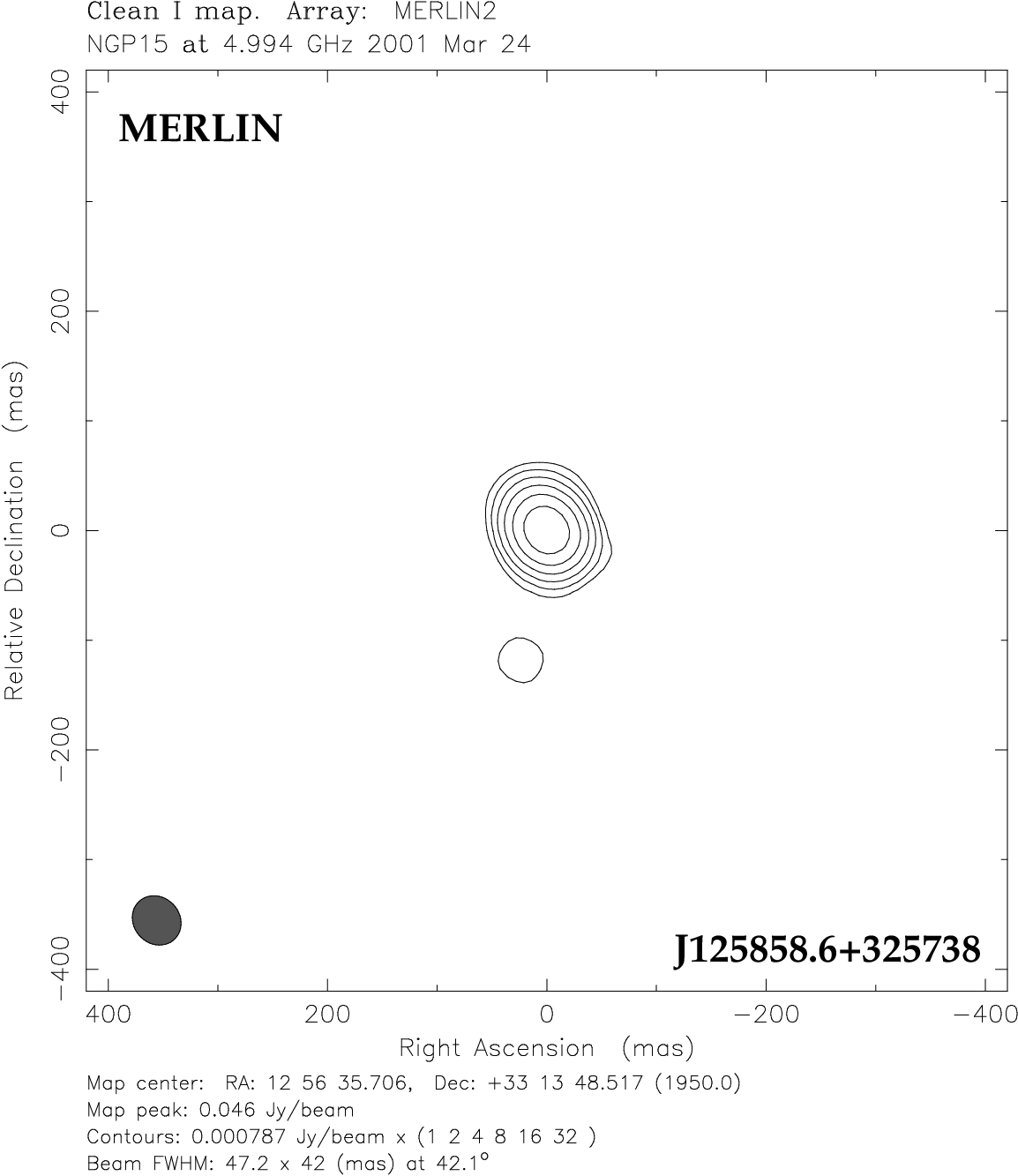}
    \includegraphics[width=7cm,bb= 32 132 583 680,clip= ]{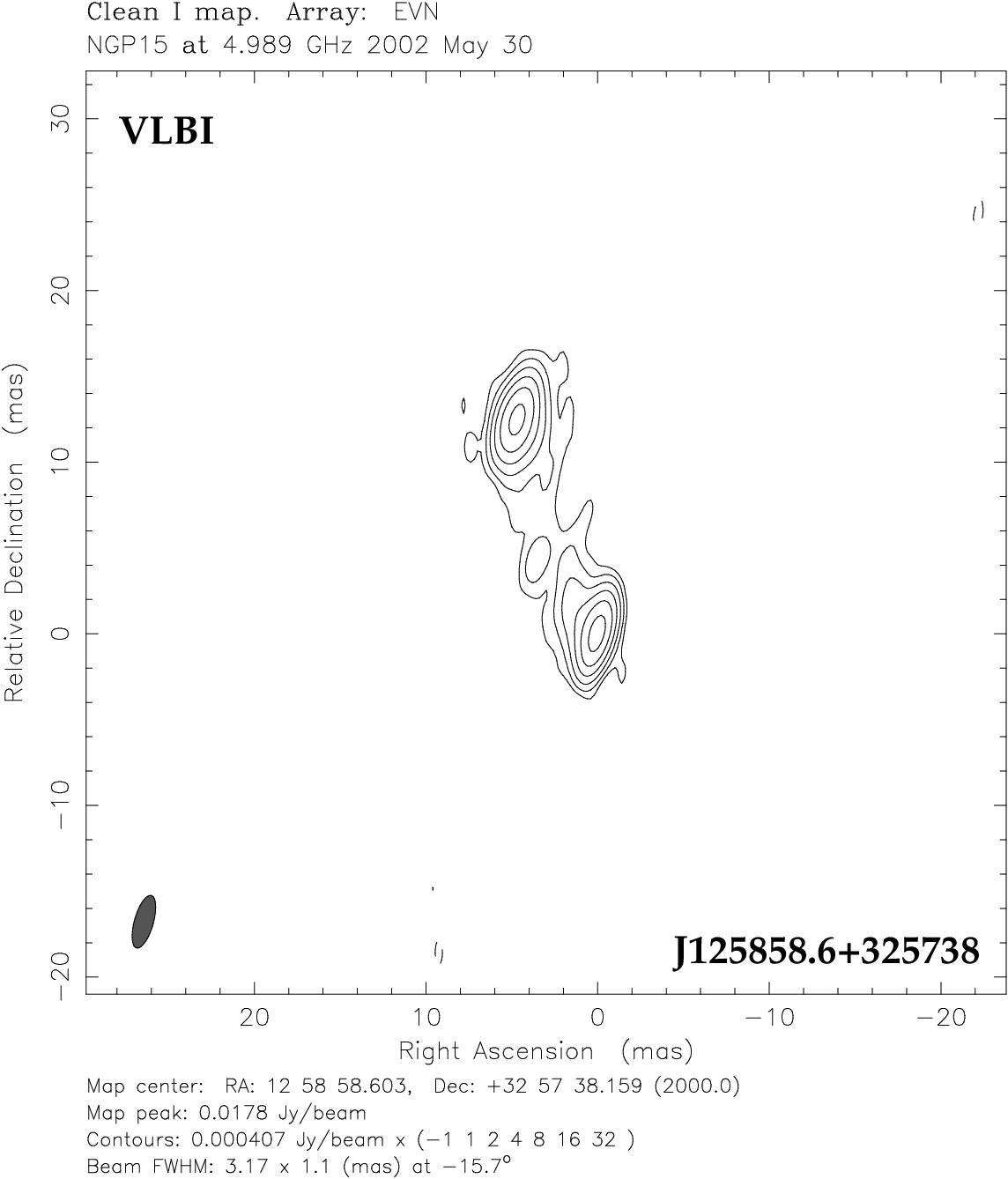}
  \caption{{\it (continued)}}
\end{figure*}

\addtocounter{figure}{-1}
\begin{figure*}
  \centering
    \includegraphics[width=7cm,bb= 32 132 588 680,clip= ]{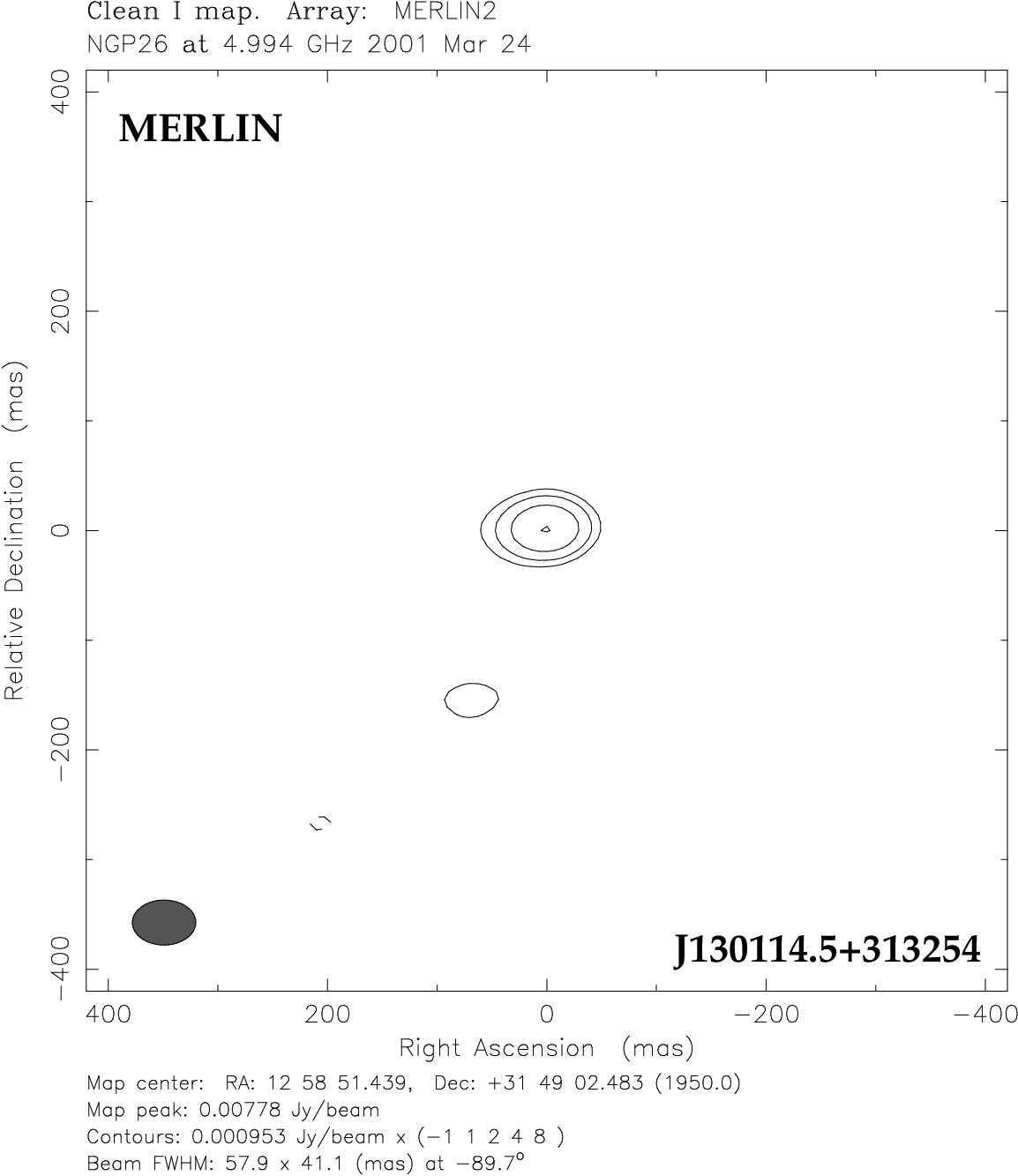}
    \includegraphics[width=7cm,bb= 32 132 583 680,clip= ]{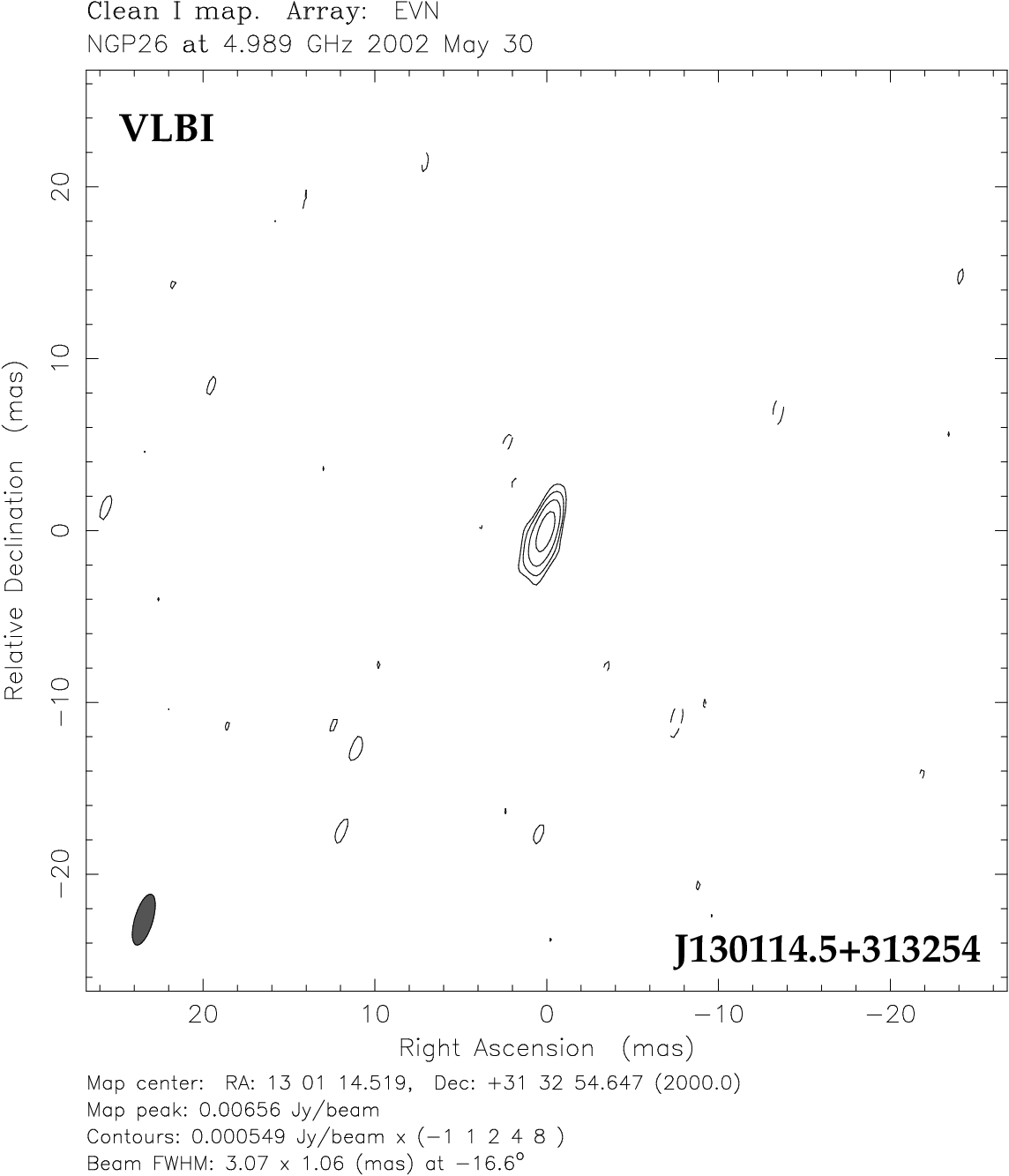}
    \includegraphics[width=7cm,bb= 32 132 588 680,clip= ]{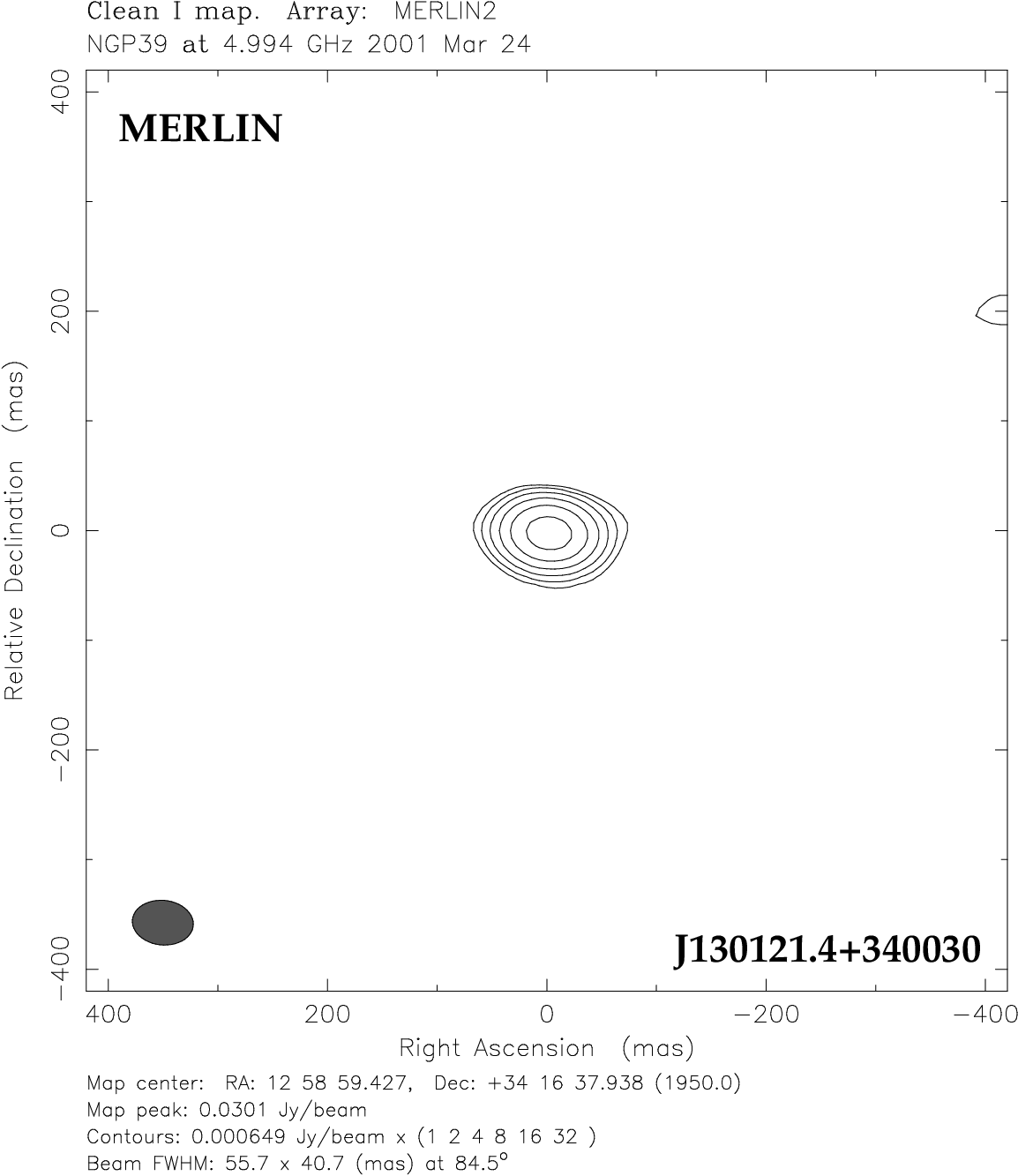}
    \includegraphics[width=7cm,bb= 32 132 583 680,clip= ]{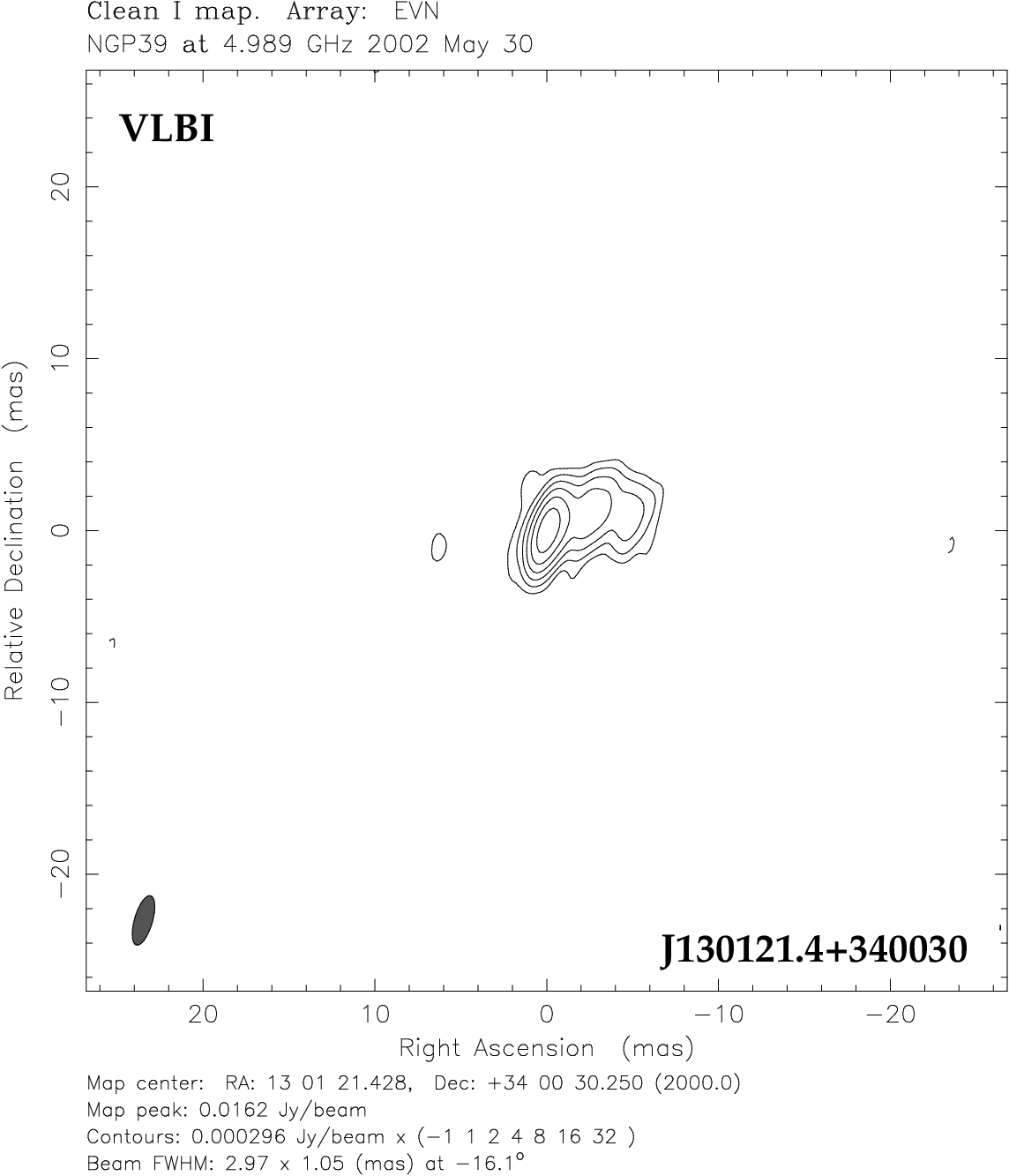}
    \includegraphics[width=7cm,bb= 32 132 588 680,clip= ]{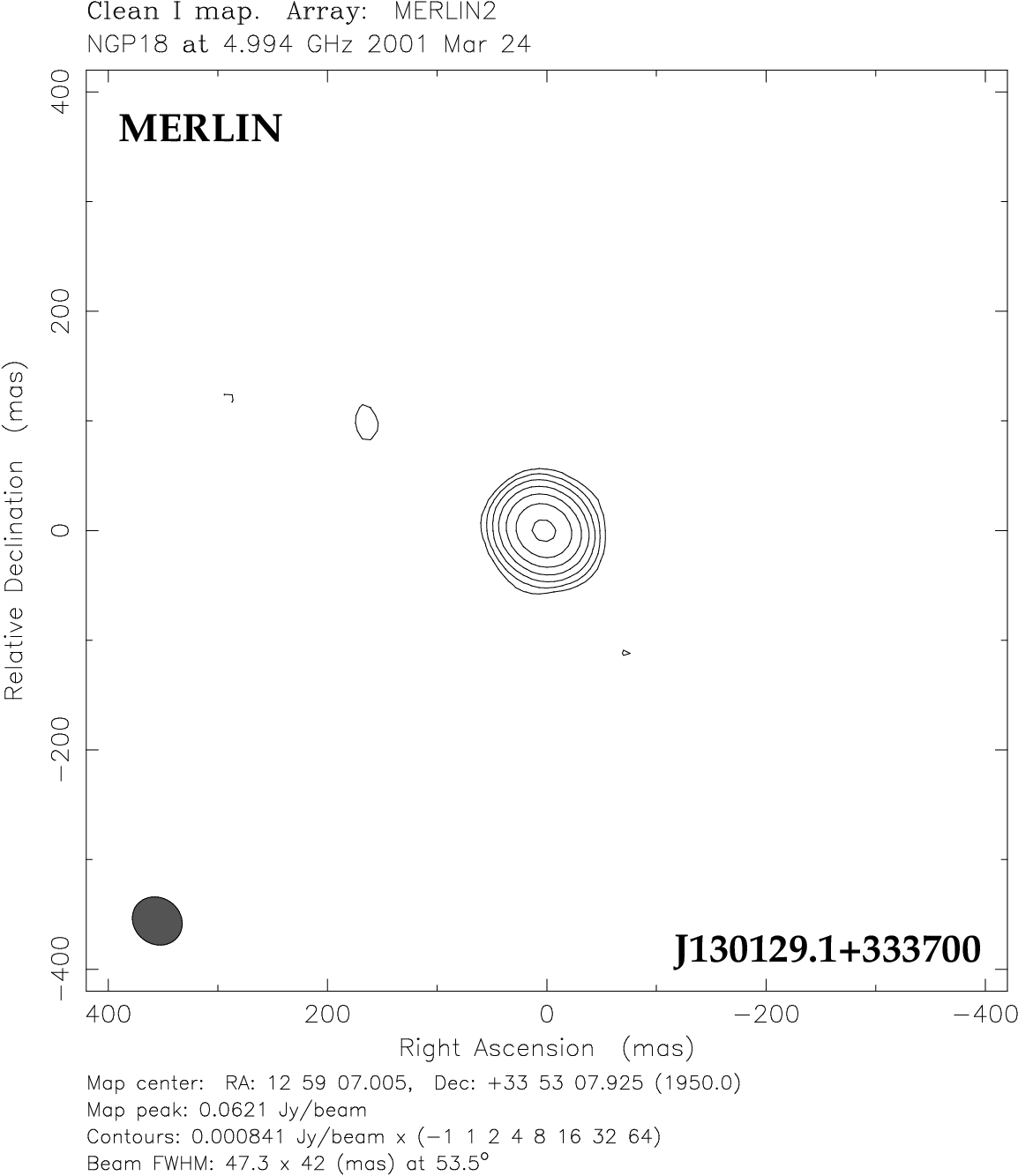}
    \includegraphics[width=7cm,bb= 32 140 583 688,clip= ]{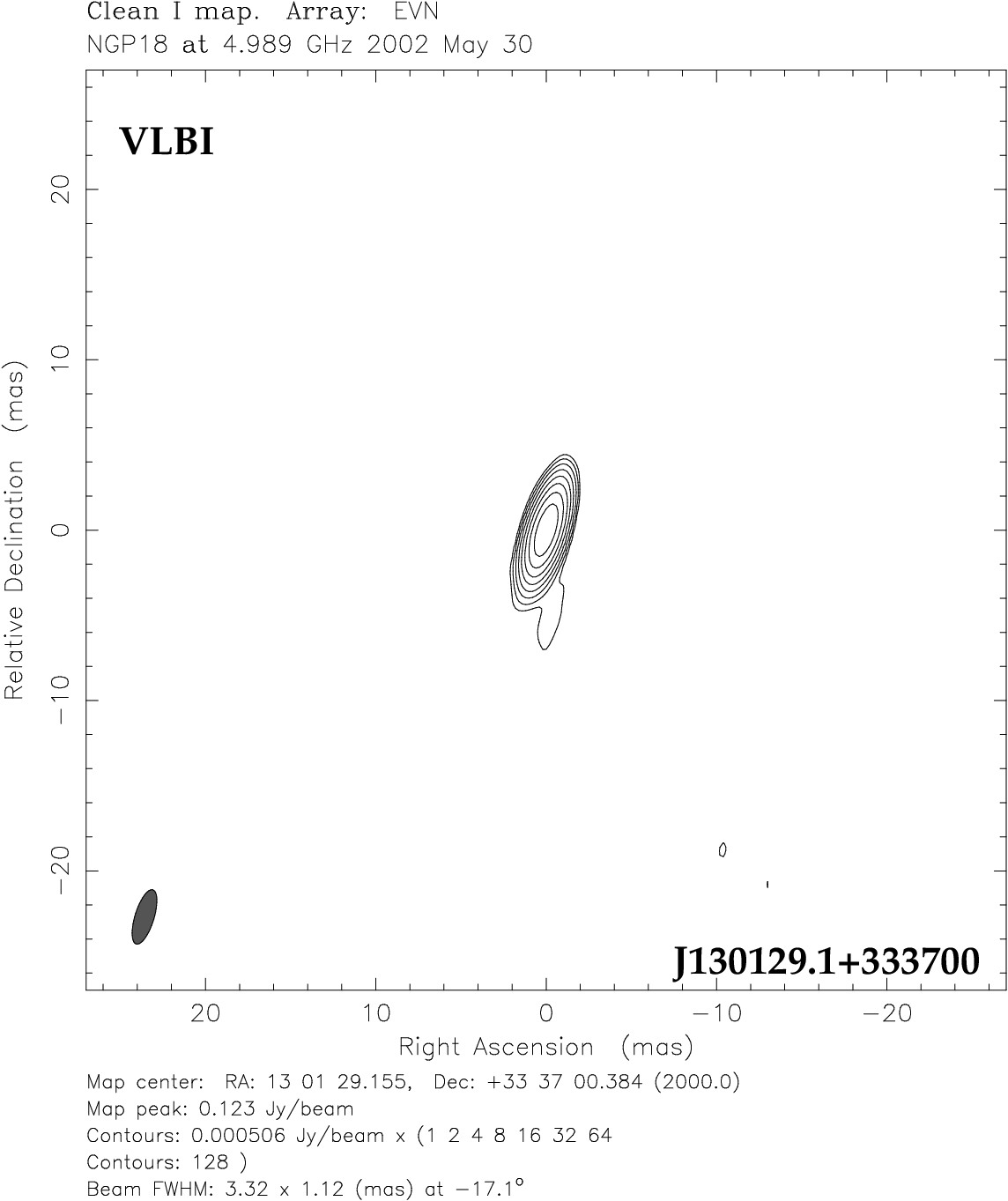}
  \caption{{\it (continued)}}
\end{figure*}

\addtocounter{figure}{-1}
\begin{figure*}
  \centering
    \includegraphics[width=7cm,bb= 32 132 588 680,clip= ]{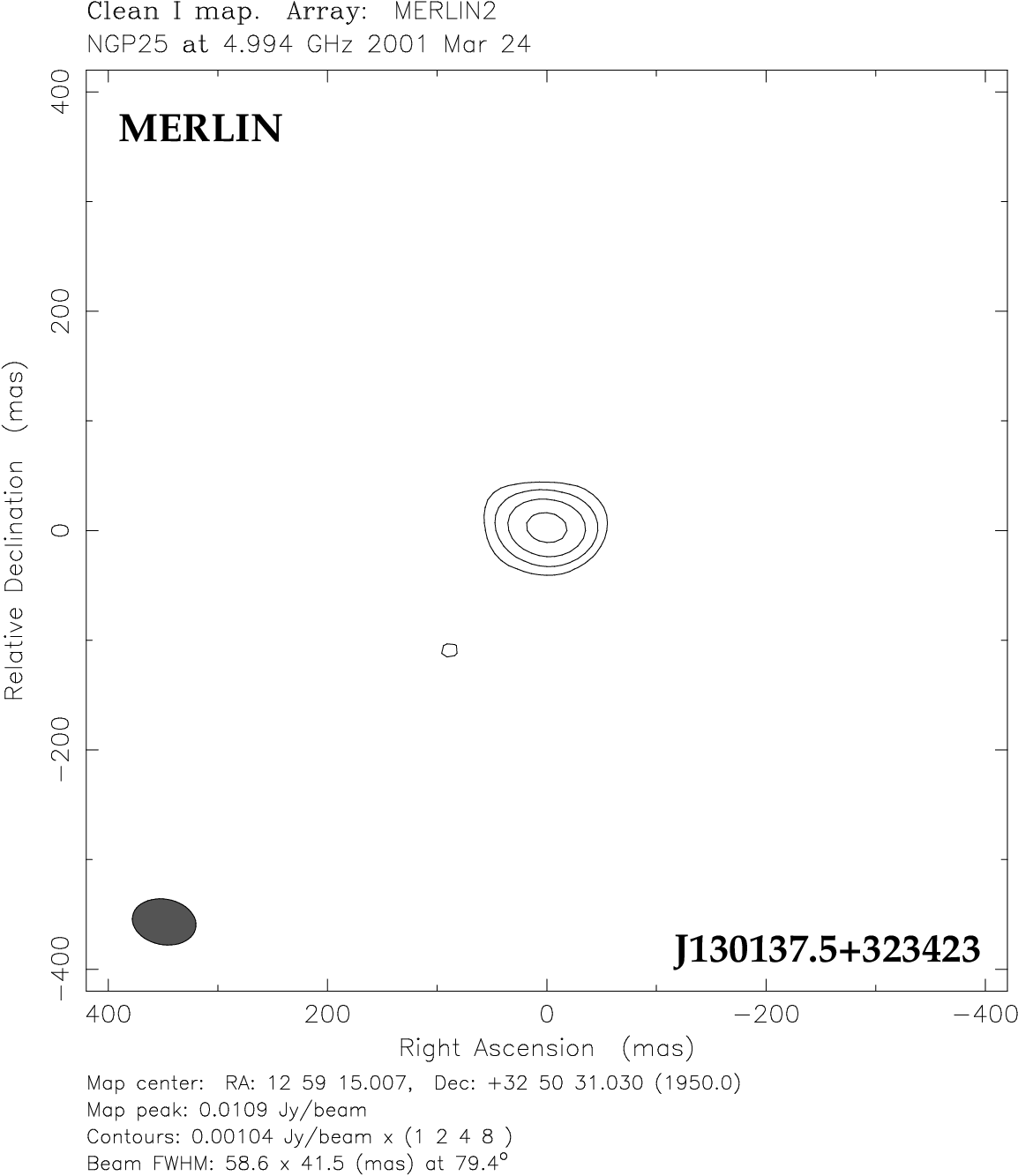}
    \includegraphics[width=7cm,bb= 32 132 583 680,clip= ]{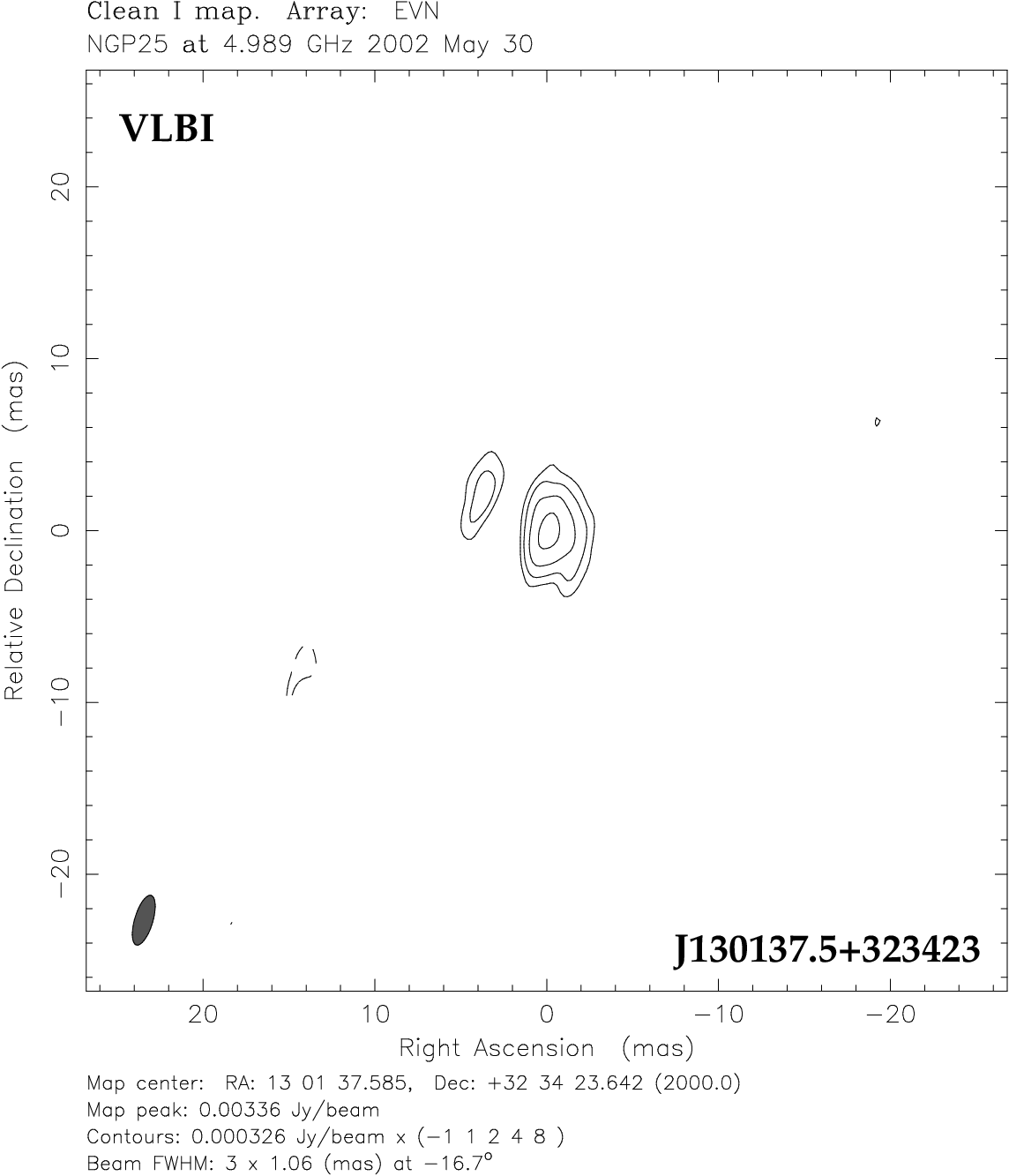}
    \includegraphics[width=7cm,bb= 32 132 588 680,clip= ]{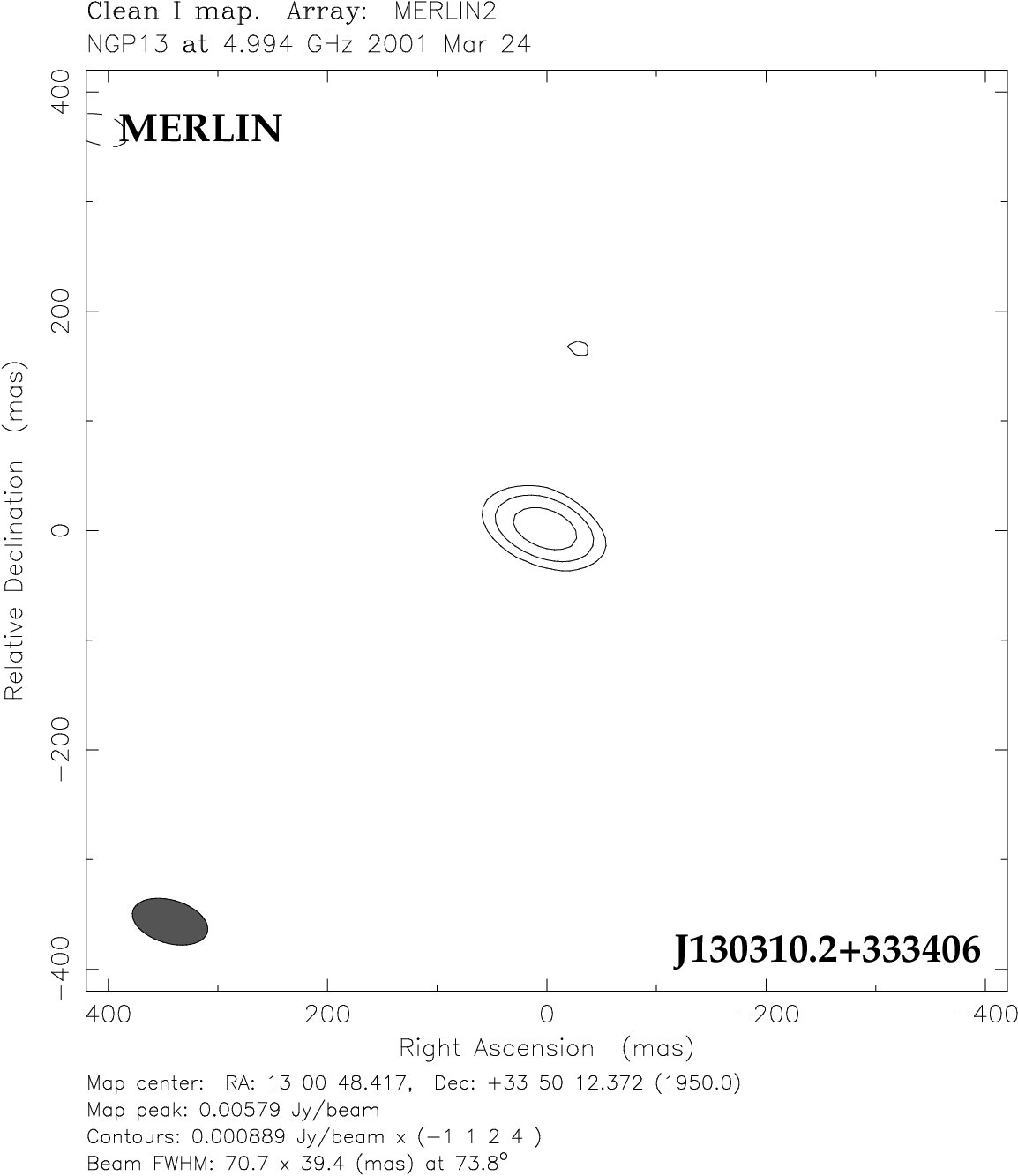}
    \includegraphics[width=7cm,bb= 32 132 583 680,clip= ]{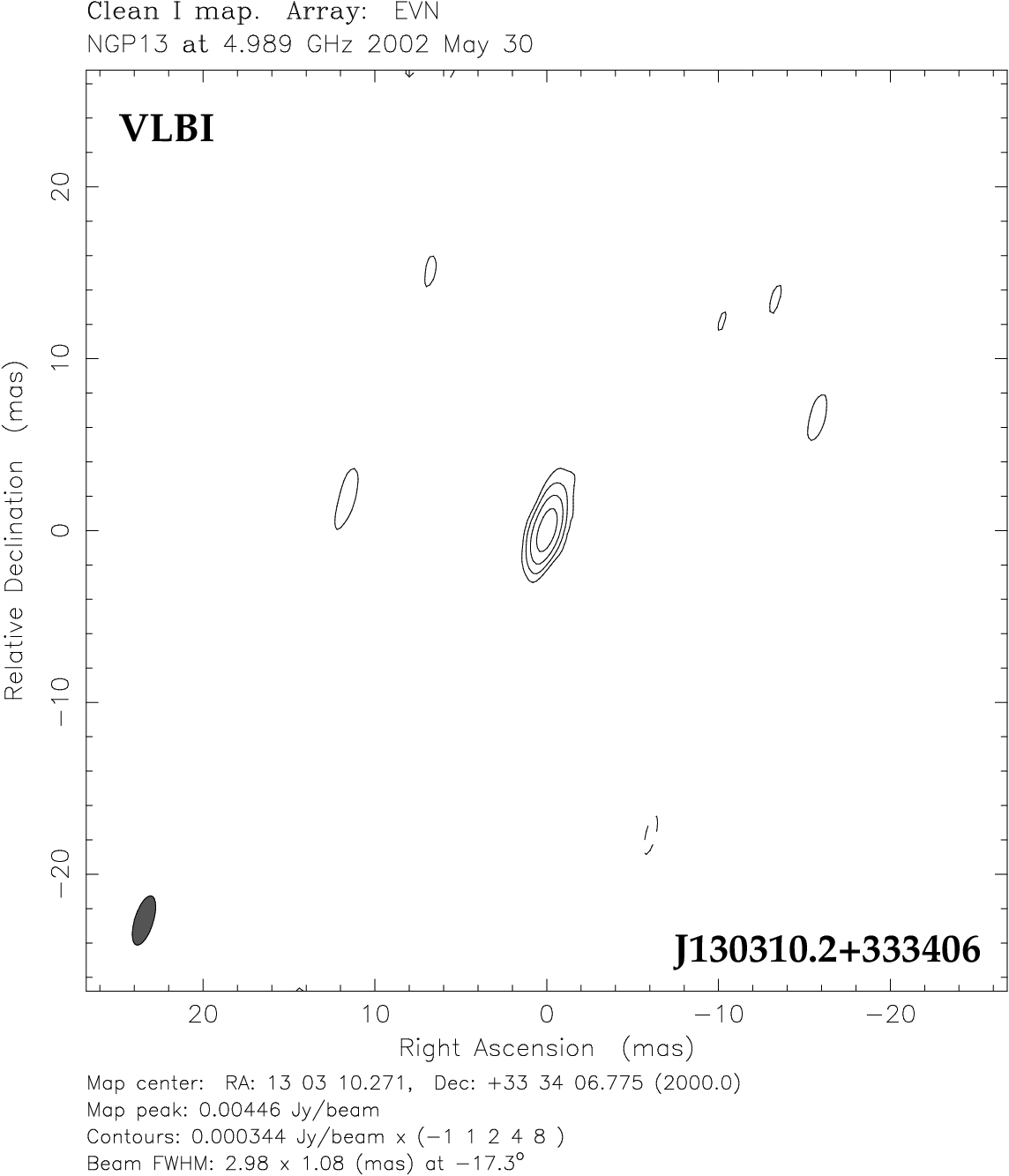}
    \includegraphics[width=7cm,bb= 32 132 588 680,clip= ]{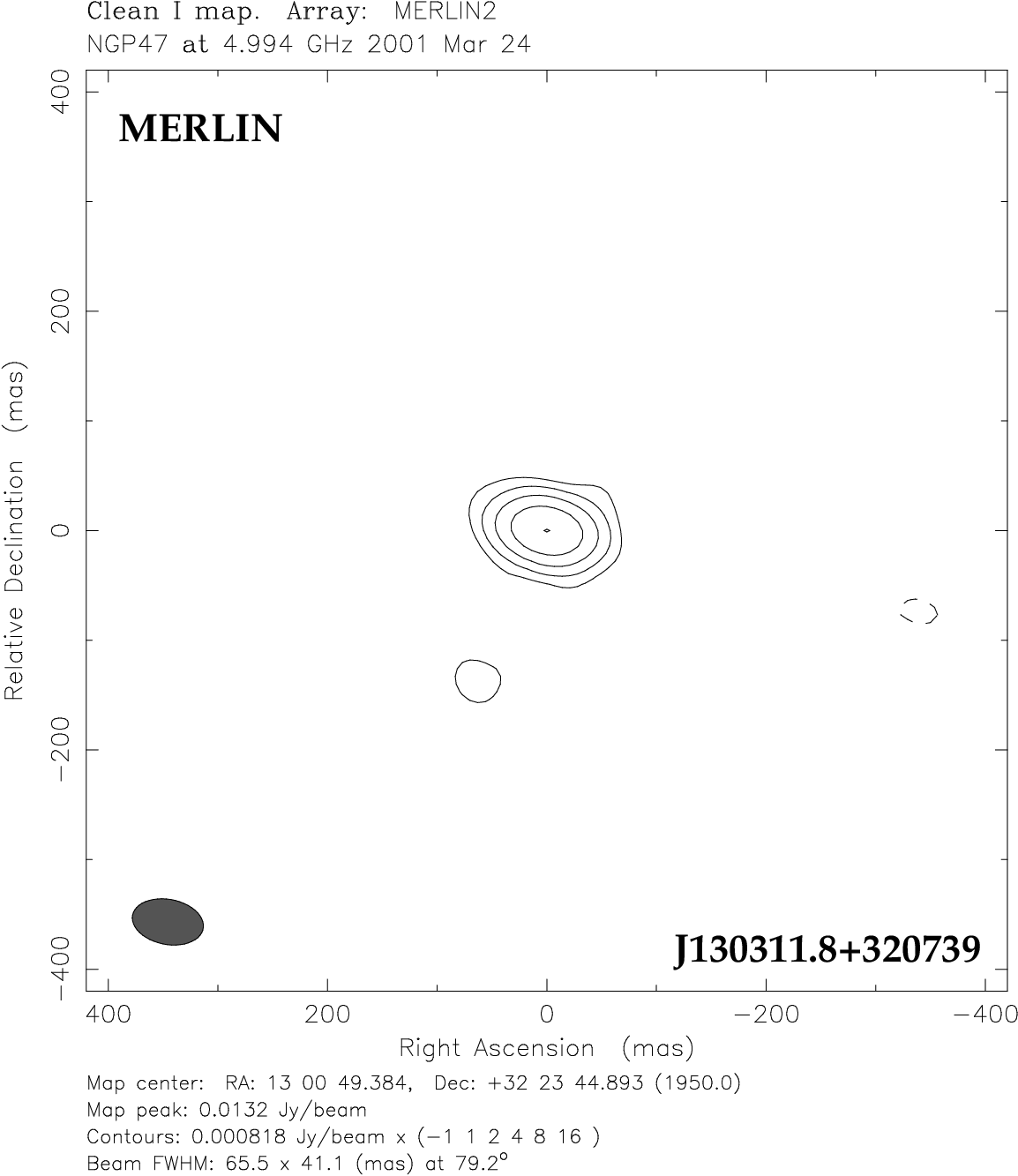}
    \includegraphics[width=7cm,bb= 32 132 583 680,clip= ]{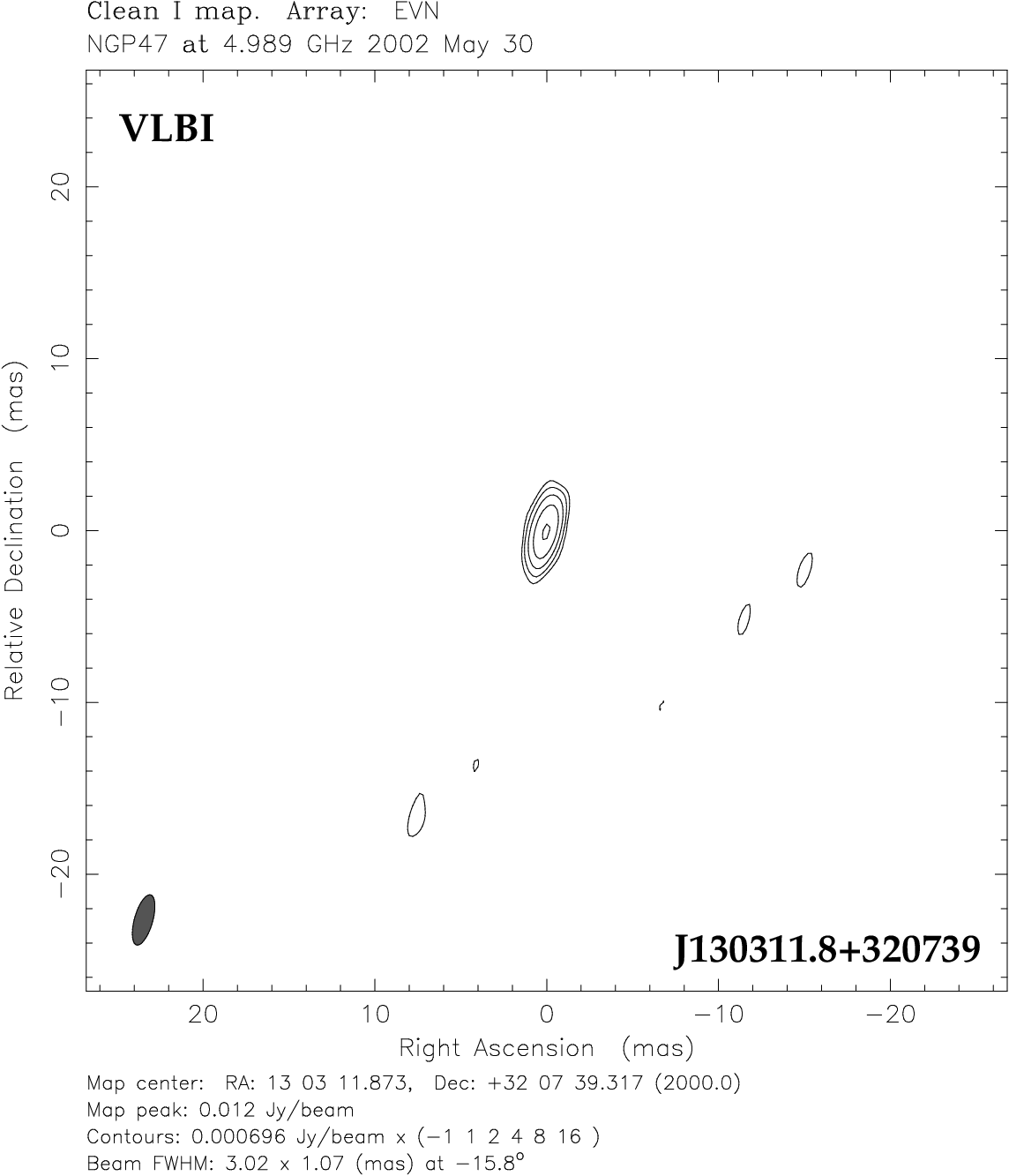}
  \caption{{\it (continued)}}
\end{figure*}

\begin{figure*}
  \centering
    \includegraphics[width=8.8cm,bb= 32 132 583 680,clip= ]{3473fig4a.ps}
    \includegraphics[width=11.8cm,angle=-90,bb= 50 150 540 700,clip= ]{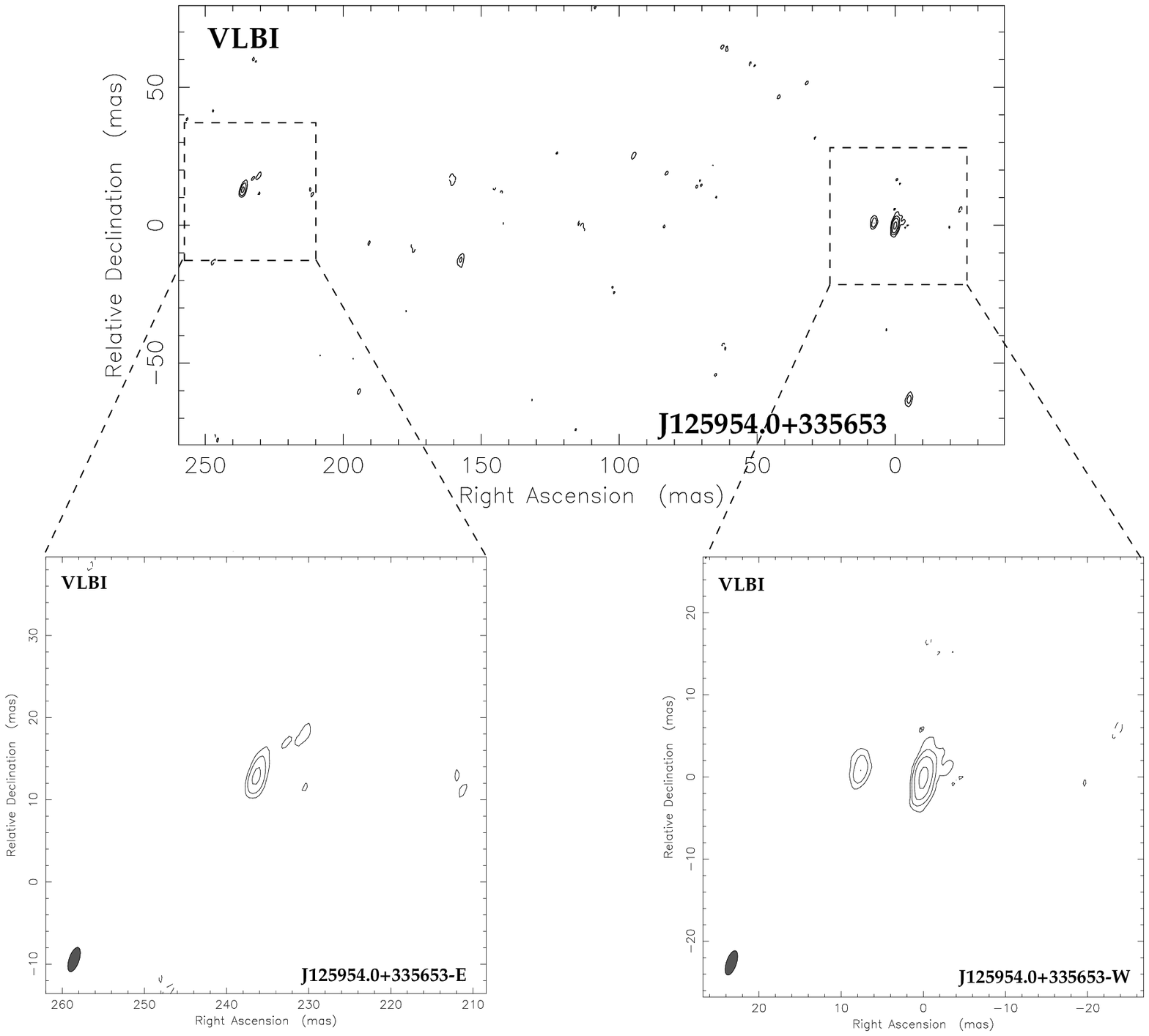}
  \caption{Naturally weighted 5-GHz MERLIN ({\it top}) and VLBI ({\it middle})
  images of J125954.0+335653. The image parameters are listed in 
Table~\ref{VLBI-maps}. The positive contour levels increase by
a factor of 2. The coordinates are relative to the VLBI brightness peak 
for 
which the absolute coordinates are given in Table~\ref{sample_ordered}. 
The restoring beam is indicated with an ellipse in the lower-left corner. 
The two panels in the bottom show an expanded view of the two brightest 
VLBI components. The peak brightness of the secondary (eastern) component 
({\it bottom left}) is 3.9 mJy/beam.}
  \label{vplot3}
\end{figure*}

\begin{table*}[b]
 \caption{MERLIN and VLBI image parameters for sources imaged with VLBI
 (Fig.~\ref{vplot1}, \ref{vplot3})}
 \label{VLBI-maps}
 \begin{tabular}{@{}lcrclr}
  \hline
  Source name & Observing & Peak brightness & Lowest contour ($3\sigma$)&
\multicolumn{2}{c}{Restoring beam} \\
              & array    & \multicolumn{2}{c}{(mJy/beam)} &
    size (mas$\times$mas) & PA ($\degr$) \\
 \hline
  J124833.1+323207 & MERLIN & 9.4 & 0.98 & $53.4 \times 37.8$ &
$57.2$ \\ 
        & VLBI & 5.2 & 0.61 & $3.16 \times 1.01$ & $-20.6$ \\
  J125240.2+331058 & MERLIN & 129.0 & 0.90 & $53.6 \times 36.8$ &
$42.8$ \\ 
        & VLBI & 74.9 & 0.51 & $3.08 \times 1.07$ & $-18.9$ \\
  J125452.4+311819 & MERLIN & 10.2 & 0.85 & $56.4 \times 36.0$ &
$36.7$ \\ 
        & VLBI & 2.4 & 0.36 & $2.90 \times 1.07$ & $-15.9$ \\
  J125755.7+313915 & MERLIN & 11.6 & 0.82 & $50.7 \times 37.4$ &
$34.0$ \\ 
        & VLBI & 9.0 & 0.50 & $2.89 \times 1.06$ & $-15.6$ \\
  J125842.2+341109 & MERLIN & 4.1 & 0.86 & $44.6 \times 44.4$ &
$-70.6$ \\ 
        & VLBI & 0.79 & 0.27 & $3.17 \times 1.08$ & $-15.2$ \\
  J125858.6+325738 & MERLIN & 46.0 & 0.79 & $47.2 \times 42.0$ &
$42.1$ \\ 
        & VLBI & 17.8 & 0.41 & $3.17 \times 1.10$ & $-15.7$ \\
  J125954.0+335653 & MERLIN & 74.1 & 0.91 & $47.3 \times 42.8$ &
$54.7$ \\ 
        & VLBI & 7.9 & 0.78 & $3.11 \times 1.22$ & $-18.8$ \\
  J130114.5+313254 & MERLIN & 7.8 & 0.95 & $57.9 \times 41.1$ &
$-89.7$ \\ 
        & VLBI & 6.6 & 0.55 & $3.07 \times 1.06$ & $-16.6$ \\
  J130121.4+340030 & MERLIN & 30.1 & 0.65 & $55.7 \times 40.7$ &
$84.5$ \\ 
        & VLBI & 16.2 & 0.30 & $2.97 \times 1.05$ & $-16.1$ \\
  J130129.1+333700 & MERLIN & 62.1 & 0.84 & $47.3 \times 42.0$ &
$53.5$ \\ 
        & VLBI & 123.2 & 0.51 & $3.32 \times 1.12$ & $-17.1$ \\
  J130137.5+323423 & MERLIN & 10.9 & 1.04 & $58.6 \times 41.5$ &
$79.4$ \\ 
        & VLBI & 3.4 & 0.33 & $3.00 \times 1.06$ & $-16.7$ \\
  J130310.2+333406 & MERLIN & 5.8 & 0.89 & $70.7 \times 39.4$ &
$73.8$ \\ 
        & VLBI & 4.5 & 0.34 & $2.98 \times 1.08$ & $-17.3$ \\
  J130311.8+320739 & MERLIN & 13.2 & 0.82 & $65.5 \times 41.1$ &
$79.2$ \\ 
        & VLBI & 12.0 & 0.70 & $3.02 \times 1.07$ & $-15.8$ \\
  \hline
  \end{tabular}
\\
Notes: Col.~1 -- source name; Col.~2 -- interferometer array name; Col.~3 -- peak
brightness at 5~GHz (mJy/beam); Col.~4 -- lowest contour level (mJy/beam) corresponding
to $3\sigma$ image noise; Col.~5 -- Gaussian restoring beam size (mas$\times$mas);
Col.~6 -- restoring beam major axis position angle ($\degr$) measured from north through east.
\end{table*}

\begin{figure*}
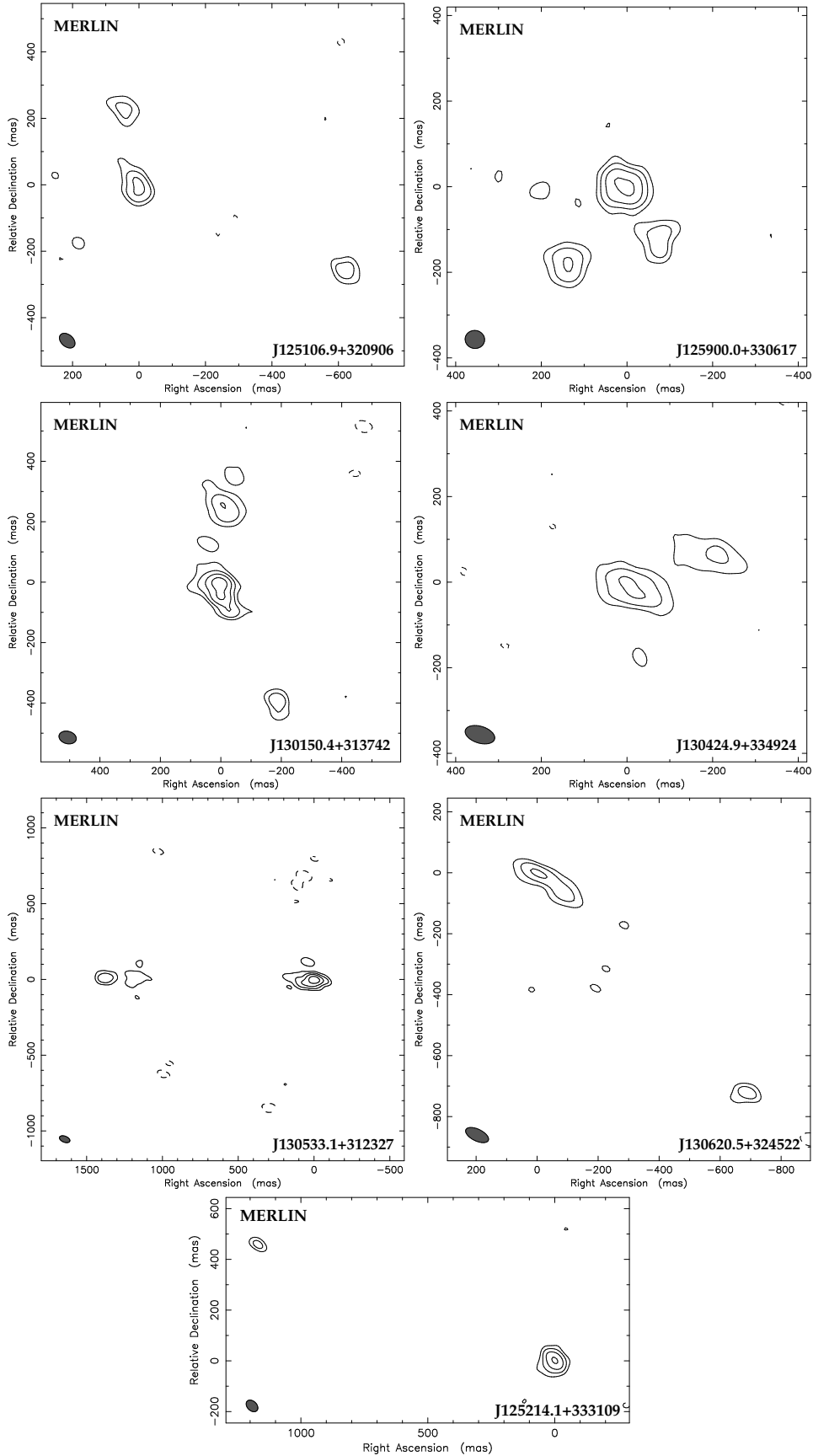

  \centering
    \includegraphics[width=6.3cm,bb= 32 132 583 680,clip=]{3473fig5a.ps}
    \includegraphics[width=6.3cm,bb= 32 132 588 680,clip=]{3473fig5b.ps}
    \includegraphics[width=6.3cm,bb= 32 132 588 680,clip=]{3473fig5c.ps}
    \includegraphics[width=6.3cm,bb= 32 132 588 680,clip=]{3473fig5d.ps}
    \includegraphics[width=6.3cm,bb= 32 132 583 680,clip=]{3473fig5e.ps}
    \includegraphics[width=6.3cm,bb= 32 132 583 680,clip=]{3473fig5f.ps}
    \includegraphics[width=7cm,bb= 32 245 583 570,clip= ]{3473fig5g.ps}
  \caption{Naturally weighted 5-GHz MERLIN images of 7 sources with resolved structure at
MERLIN angular scales but not detected or marginally detected with VLBI. 
The image parameters are listed in
Table~\ref{MERLIN-maps}. The positive contour levels increase by a factor of 2.
The coordinates are relative to the MERLIN brightness peak for which the 
absolute coordinates are given in Table~\ref{sample_ordered}. The 
restoring beam is indicated with an ellipse in the lower-left corner.}
   \label{mplot1}
\end{figure*}

\begin{table*}
 \caption{MERLIN image parameters for sources with resolved structure at
MERLIN angular scales but not detected with VLBI (Fig.~\ref{mplot1})}
 \label{MERLIN-maps}
 \begin{tabular}{@{}lrclr}
  \hline
  Source name & Peak brightness & Lowest contour ($3\sigma$)&
\multicolumn{2}{c}{Restoring beam} \\
              & \multicolumn{2}{c}{(mJy/beam)} &
    size (mas$\times$mas) & PA ($\degr$)\\
 \hline
 J125106.9+320906 & 5.4 & 1.11 & $53.5 \times 36.1$ & $49.1$ \\
 J125214.1+333109 & 10.1 & 1.14 & $53.4 \times 37.2$ & $45.8$ \\
 J125900.0+330617& 11.4 & 1.25 & $45.6 \times 42.0$ & $81.2$ \\
 J130150.4+313742 & 12.1 & 0.97 & $58.3 \times 41.1$ & $75.2$ \\
 J130424.9+334924 & 6.7 & 1.44 & $71.4 \times 39.2$ & $72.7$ \\
 J130533.1+312327 & 15.9 & 1.20 & $72.5 \times 40.0$ & $68.6$ \\
 J130620.5+324522 & 4.3 & 0.92 & $81.6 \times 38.7$ & $64.2$ \\
  \hline
  \end{tabular}
\\
Notes: Col.~1 -- source name; Col.~2 -- peak brightness at 5~GHz (mJy/beam);
Col.~3 -- lowest contour level (mJy/beam) corresponding to $3\sigma$ image noise;
Col.~4 -- Gaussian restoring beam size (mas$\times$mas); Col.~5 -- restoring beam
major axis position angle ($\degr$) measured from north through east.
\end{table*}

\end{document}